\documentclass[fleqn,usenatbib,useAMS]{mnras}

\usepackage{graphicx}	
\usepackage{amsmath}	
\usepackage{multicol}   
\usepackage{pdflscape}	
\usepackage{multirow}
\usepackage{booktabs}
\usepackage{siunitx}
\usepackage{float}

\usepackage[T1]{fontenc}
\usepackage{ae,aecompl}

\usepackage{newtxtext,newtxmath}

\title[Solar Galactocentric distance from Gaia DR3]
{The solar Galactocentric distance and local kinematic parameters from Gaia DR3 using Bottlinger's equations}

\author[D. V. Odynets et al.]{
D. V. Odynets, $^{1}$ \thanks{Contact e-mail: odynets.d.v@gmail.com (DVO)}
V. S. Akhmetov, $^{1,2,3}$ \thanks{Contact e-mail: akhmetovvs@gmail.com (VSA)}
S. I. Denyshchenko $^{1}$ 
\\
$^{1}$Institute of Astronomy of V.N. Karazin Kharkiv National University, Svobody sq. 4, 61022, Kharkiv, Ukraine \\
$^{2}$INAF--Osservatorio Astrofisico di Torino, Via Osservatorio 20, Pino Torinese, Turin, I-10025, Italy \\
$^{3}$Main Astronomical Observatory of the NAS of Ukraine, 27 Akademika Zabolotnoho St., Kyiv 03143, Ukraine
}

\date{Accepted XXX. Received YYY; in original form ZZZ}
\pubyear{\the\year{}}

\begin{document}
\label{firstpage}
\pagerange{\pageref{firstpage}--\pageref{lastpage}}

\maketitle

\begin{abstract}
The solar Galactocentric distance, $R0$, together with the local kinematic parameters of the Milky Way, provides fundamental constraints on the Galactic structure and rotation. Their accurate determination has become possible with the hight astrometric precision of Gaia DR3 catalogue.
We determine the Galactocentric distance $R_0$, the Solar peculiar motion components $(u_0, v_0, w_0)$, and the local angular velocity of Galactic rotation, $\omega_0$, using a self-consistent approach based on Bottlinger's equations within a nonlinear least-squares framework.
Bottlinger's equations were applied to approximately 13 million luminous giant stars ($M_G < 4$) selected from Gaia DR3. The analysis was carried out for two subsamples: a thin disk dominated sample ($|z| < 0.5\,\mathrm{kpc}$) and a mixed thin and thick disk sample ($|z| < 1.0\,\mathrm{kpc}$). The kinematic parameters were estimated using a nonlinear generalized least-squares method with the Levenberg--Marquardt optimization. The full Gaia covariance matrices were incorporated through Cholesky whitening, and parameter uncertainties were estimated from Monte Carlo realizations.
We obtained $R_0 = 8.165 \pm 0.024\,\mathrm{kpc}$ and a local angular velocity of Galactic rotation $\omega_0 = 28.477 \pm 0.002\,\mathrm{km\,s^{-1}\,kpc^{-1}}$ for the thin disk dominated sample. The Solar peculiar motion components are $(u_0, v_0, w_0) = (9.261 \pm 0.006,\,16.593 \pm 0.008,\,7.686 \pm 0.008)\,\mathrm{km\,s^{-1}}$. Including mixed thin and thick disk stars resulted in a systematic shift in $R_0$, yielding $R_0 = 8.333 \pm 0.022\,\mathrm{kpc}$, while the Solar peculiar motion and local angular velocity remain consistent within uncertainties.
The Galactic kinematic parameters derived are consistent with recent independent determinations. The Solar peculiar motion parameters and the local angular velocity remain stable for different vertical selections, the recovered Galactocentric distance exhibits a systematic dependence on the adopted stellar sample, indicating that the vertical structure of the tracer population should be taken into account when determining $R_0$.
\end{abstract}


\section{Introduction}
\label{sec:intro}

Understanding the structure and dynamical evolution of the Milky Way requires accurate knowledge of its fundamental parameters \citep{BlandHawthorn2016}. Among the most important are the Galactocentric distance $R_0$, the components of the Solar peculiar motion $(u_0, v_0, w_0)$, and the local angular velocity of Galactic rotation $\omega_0$. Together, these quantities provide the fundamental reference frame for studies of Galactic kinematics and the Galactic rotation curve \citep{Reid2014, McMillan2017, Dmytrenko2023}. These quantities cannot be measured directly but must instead be inferred by fitting kinematic models to observational data. Their determination is therefore inherently model-dependent and sensitive to the adopted tracer population, sample selection, and underlying model assumptions \citep{BinneyTremaine2008,McMillan2017,Fedorov2026}. Although the high astrometric precision of modern surveys has substantially reduced statistical uncertainties, systematic effects, the complexity of the Galactic velocity field, and correlations among simultaneously estimated parameters remain major challenges in obtaining a robust and self-consistent kinematic solution \citep{BlandHawthorn2016, Lindegren2021, Brown2021}.

The advent of the Gaia mission has fundamentally transformed studies of Galactic structure and kinematics by providing astrometric measurements of unprecedented precision and homogeneity for more than one billion stars \citep{Gaia2016, Gaia2023}. In addition to positions, parallaxes, and proper motions, Gaia DR3 provides radial velocity measurements for more than 30 million stars, enabling large-scale studies of the six-dimensional phase-space structure of the Milky Way \citep{Katz2023}. These advances have enabled the application of both classical and newly developed kinematic methods to vastly larger stellar samples, substantially improving constraints on Galactic kinematic parameters and the structure and dynamics of the Milky Way \citep{Gaia2023, Brown2021, Akhmetov2024}.

The solar Galactocentric distance, $R_0$, has been investigated for more than a century. One of the first systematic estimates was obtained by \citet{Shapley1918}, who inferred the position of the Galactic centre from the spatial distribution of globular clusters and estimated $R_0$ to be between approximately $13$ and $25,\mathrm{kpc}$. Subsequent studies based on globular clusters \citep[$R_0\simeq8.5,\mathrm{kpc}$]{Harris1976} and classical Cepheids \citep[$R_0\simeq7.7,\mathrm{kpc}$]{Majaess2009} progressively refined the Galactic distance scale. Advances in observational techniques, astrometric precision, and Galactic dynamical modelling have since led to a broad range of complementary methods for determining $R_0$ and other key Galactic kinematic parameters.

At present, the fundamental parameters describing the Galactic structure and kinematics are determined using several complementary observational approaches. Long-term astrometric and spectroscopic monitoring of stars orbiting the supermassive black hole Sagittarius~A* has provided some of the most direct geometric estimates of the Galactocentric distance and the mass of the central black hole. Early orbital analyses by \citet{Gillessen2017} were subsequently refined by \citet{Do2019}. Infrared interferometric observations by the GRAVITY Collaboration further improved the astrometric precision and tightened the constraints on both $R_0$ and the mass of the central black hole \citep{Gravity2018, Gravity2021}.

Very Long Baseline Interferometry (VLBI) observations of masers associated with high-mass star-forming regions provide additional constraints to Galactic parameters. Early studies by \citet{Reid2014} were subsequently refined using larger maser samples \citep{Reid2019}, while independent observations with the VERA array provided an additional geometric constraint on the Galactic distance scale \citep{VERA2020}. By combining trigonometric parallaxes, proper motions, and radial velocities of maser sources distributed throughout the Galactic disc, these studies simultaneously constrain the Galactic rotation curve and the large-scale Galactic kinematics.

Standard candles provide another independent approach to estimating the Galactocentric distance. RR Lyrae stars were employed by \citet{Dekany2013} and \citet{Pietrukowicz2015} to derive estimates of $R_0$, while \citet{Nataf2013} used red clump giants to constrain the Galactic distance scale and the three-dimensional structure of the Galactic bulge. Owing to their well-calibrated luminosities, these stellar tracers provide an important independent check on geometric determinations of $R_0$.

Complementary constraints are provided by kinematic and dynamical approaches that model the large-scale stellar velocity field and allow for a self-consistent determination of the key Galactic kinematic parameters. These approaches include axisymmetric rotation-curve modelling \citep{Bovy2012,McMillan2017}, stellar kinematic analyses \citep{Schonrich2010,Xu2018}, studies of Galactic bar kinematics \citep{Leung2022}, and investigations of the local velocity field using different kinematic models \citep{Fedorov2021,Bobylev2023,Akhmetov2024}.

Current observational estimates place the Galactocentric distance of the Sun in the range $R_0 \simeq 8.1$--$8.3$~kpc, although differences remain among different methods and tracer populations \citep{BlandHawthorn2016, deGrijs2016}. The components of the Solar peculiar motion are typically found within the ranges $u_0 \simeq 8$--$11$, $v_0 \simeq 10$--$18$, and $w_0 \simeq 7$--$9~\mathrm{km\,s^{-1}}$, while the local angular velocity of Galactic rotation is generally estimated to be $\omega_0 \simeq 26$--$30~\mathrm{km\,s^{-1}\,kpc^{-1}}$ \citep{Schonrich2010, Schonrich2012, McMillan2017, Reid2019}.

Bottlinger's equations provide a description of the local stellar velocity field by relating the observed radial velocities and proper motions of stars to the Solar peculiar motion and the differential rotation of the Galaxy. Within this framework, the solar Galactocentric distance, the components of the Solar peculiar motion, the local angular velocity of Galactic rotation and its radial derivatives, as well as a parameter describing the radial expansion/contraction of the stellar system, can be determined simultaneously from stellar kinematic observations \citep{Ogorodnikov1965,Kulikovsky1985}. Estimating these parameters within a common kinematic model allows their mutual correlations to be accounted for consistently and makes the method well suited to the homogeneous six-dimensional Gaia DR3 data set. In this work, we apply this approach to determine the fundamental Galactic parameters and investigate their dependence on the adopted spatial extent of the stellar sample. Parameter uncertainties are estimated using Monte Carlo simulations.

This paper is organized as follows. Section~\ref{sec:data} describes the Gaia DR3 data set, the adopted selection criteria, and the construction of the stellar samples. Section~\ref{sec:met} presents the Bottlinger kinematic model and the parameter estimation procedure. The resulting kinematic parameters are presented in Section~\ref{sec:res}. Section~\ref{sec:dis} discusses the results in comparison with previous determinations from the literature, examines the influence of the adopted stellar population on the recovered parameters, and describes the application of the method to Gaia-like mock catalogues. Finally, the main conclusions are summarized in Section~\ref{sec:con}.

\section{Data and sample}
\label{sec:data}

We selected stars from Gaia DR3 \citep{Prusti2016,Vallenari2023} with full six-dimensional phase-space information, that is, stars with five-parameter astrometric solutions (positions, parallaxes, and proper motions) combined with measured radial velocities. The initial sample contains approximately $33$ million stars.

Following the approach adopted in previous Gaia-based kinematic studies \citep{Akhmetov2024}, we applied a set of selection criteria designed to ensure reliable astrometric measurements, accurate distance estimates, and a homogeneous tracer population suitable for Galactic kinematic analysis. The individual criteria and their physical motivation are described below.

The condition $\mathrm{RUWE}<1.4$ excludes sources with potentially problematic astrometric solutions and reduces contamination from sources affected by unresolved binarity or other unmodelled astrometric effects \citep{Lindegren2018}. The parallaxes were corrected for the Gaia DR3 global parallax zero point following the prescription of \citet{Lindegren2021}. The corrected parallaxes were then used consistently in all subsequent distance estimates and kinematic calculations. Requiring $\varpi/\sigma_{\varpi}>5$ limits the relative parallax uncertainty to below $20\%$. For these nearby stars, with $r<5~\mathrm{kpc}$, the corrected parallaxes were sufficiently precise to allow heliocentric distances to be estimated using the inverse-parallax relation, $r=1/\varpi$.

To ensure consistency with the assumptions of axisymmetric and nearly circular Galactic rotation adopted in the Bottlinger equations, we retained only stars with orbital eccentricities $ecc < 0.2$, using catalogue of the Galactic orbital parameters published by \citet{Palicio2023}. This criterion excludes stars on highly eccentric orbits and therefore improves the applicability of the adopted kinematic approximation.

\begin{figure}
\centering
\includegraphics[width=0.9\columnwidth]{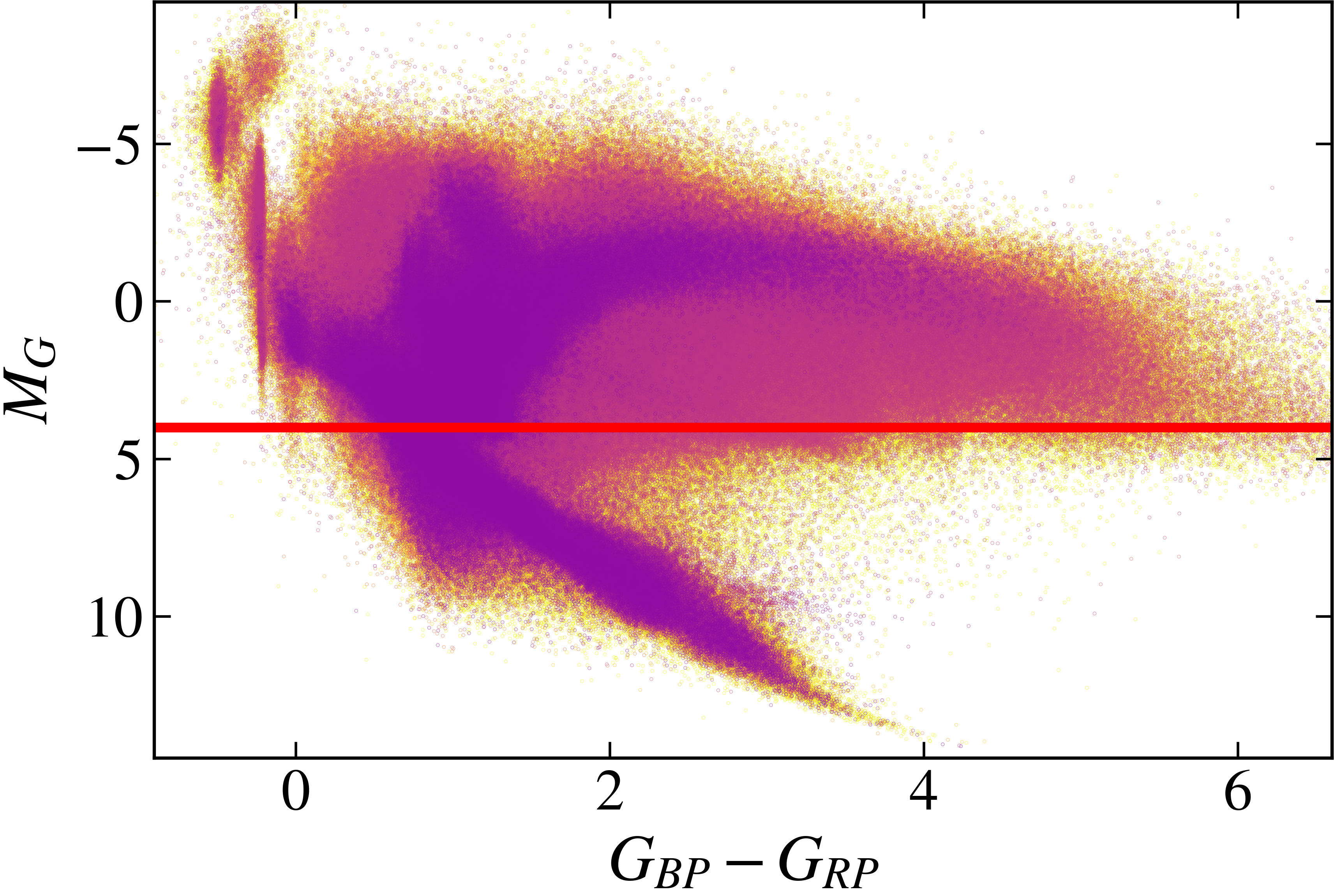}
\caption{Colour--magnitude diagram of the selected Gaia DR3 sample. The horizontal line at $M_G=4$ marks the adopted luminosity threshold used to select luminous stellar tracers, including upper main-sequence stars, subgiants, and giants.}
\label{fig:CMD}
\end{figure}

To obtain a spatially extended and approximately homogeneous tracer population, we selected only luminous stars with extinction-corrected absolute magnitudes $M_G<4$. The absolute magnitudes were computed using the Gaia DR3 extinction estimate $A_G$ together with the conservative parallax estimate $\varpi+2.32\sigma_\varpi$, following \citet{RuizDern2018}. This criterion reduces contamination from the numerous low-luminosity main-sequence stars while retaining intrinsically bright tracers, including upper main-sequence stars, subgiants, and giants. The colour--magnitude diagram in Fig.~\ref{fig:CMD} illustrates the distribution of the selected stellar tracers and the adopted luminosity cut.

The final selection is defined by the following conditions:
\begin{align} 
& \mathrm{RUWE} < 1.4, \notag \\ 
& \mathrm{non\_single\_star}=0, \notag \\ 
& \frac{\varpi}{\sigma_{\varpi}}>5, \notag \\ 
& m_G + 5 + 5\log_{10}\left(\frac{\varpi+2.32\,\sigma_\varpi}{1000}\right)-A_G < 4, \notag\\
& ecc < 0.2. \notag
\end{align}

After applying these selection criteria, the final sample contains approximately $13$ million stars.

To investigate the influence of the adopted vertical extent of the stellar sample on the derived kinematic parameters, we considered two subsamples defined by their distance from the Galactic mid-plane, $|z|<0.5$ and $|z|<1.0$ kpc. These limits were chosen to probe the sensitivity of the kinematic solution to the increasing contribution of the thick-disc population. The choice is motivated by the two-component vertical structure of the Galactic disc, with the thin- and thick-disc components having characteristic scale heights of approximately $300$ and $900$ pc, respectively \citep{Robin2003,BlandHawthorn2016}. The $|z|<0.5$ kpc sample is expected to be strongly dominated by thin-disc stars, whereas the extended $|z|<1.0$ kpc sample contains a larger fraction of thick-disc stars while remaining dominated by the thin-disc population. Comparing the corresponding solutions allows us to assess the sensitivity of the derived kinematic parameters to changes in the vertical composition of the tracer population.

\section{Methods}
\label{sec:met}

In this work, instead of the general three-dimensional kinematic model of Ogorodnikov--Milne \citep{Ogorodnikov1965}, we adopt the classical Bottlinger model of Galactic rotation \citep{Kulikovsky1985}. In this formulation, stars are assumed to move on circular orbits around the Galactic centre in planes parallel to the middle Galactic plane.

The kinematic equations are formulated in the Galactic coordinate system, where $l$ and $b$ denote the Galactic longitude and latitude, respectively, and $r$ is the heliocentric distance to a star, calculated from the Gaia DR3 parallax as $r=1/\varpi$. The observed heliocentric velocity components are denoted by $V_r$, $V_l = kr\mu_l$, and $V_b = kr\mu_b$, where $\mu_l$ and $\mu_b$ are the proper-motion components in Galactic longitude and latitude, respectively, and $k=4.74047$ is the conversion factor from $ mas\,yr^{-1}\,kpc$ to $km\,s^{-1}$.

The contribution of the Solar peculiar motion to the observed heliocentric velocities is given by

\begin{align}
\label{eq:Vsun}
  & \Delta V_{r_0} = (u_0 \cos b \cos l + v_0 \cos b \sin l + w_0 \sin b), \notag \\
  & \Delta V_{l_0} = (v_0 \cos l - u_0\sin l), \\
  & \Delta V_{b_0} = (w_0 \cos b - u_0 \sin b \cos l - v_0 \sin b \sin l). \notag
\end{align}

Here $u_0$, $v_0$, and $w_0$ are the components of the Solar peculiar velocity directed toward the Galactic center, in the direction of Galactic rotation, and toward the North Galactic Pole, respectively.

After correcting for the Solar peculiar motion, the remaining velocity field is described by the Bottlinger equations:

\begin{align}
\label{eq:bottlinger}
  & V_r + \Delta V_{r_0} = R_0 (\omega - \omega_0) \sin l \cos b, \notag \\
  & V_l + \Delta V_{l_0} = (R_0 \cos l - r \cos b)(\omega - \omega_0), \\
  & V_b + \Delta V_{b_0} = -R_0 (\omega - \omega_0)\sin l \sin b. \notag
\end{align}

From the geometry of the Sun, the Galactic center, and the star, the Galactocentric distance $R$ is obtained from the law of cosines:

\begin{align}
\label{eq:R}
R^2 = R^2_0 + r^2\cos^2 b - 2R_0r\cos\,b\cos l.
\end{align}

Since the Galactic angular velocity is a function of the Galactocentric distance, it is expanded in a Taylor series about the Solar position:

\begin{align}
\label{eq:omega}
(\omega - \omega_0) \approx \omega^\prime_0 ( R - R_0) + \frac {1}{2!} \omega^{\prime\prime}_0 ( R - R_0)^2,
\end{align}
where $\omega_0$ denotes the Galactic angular velocity at the Solar Galactocentric distance $R_0$, while $\omega^\prime_0$ and $\omega^{\prime\prime}_0$ are its first and second derivatives with respect to $R$.

Substituting Eqs.~(\ref{eq:Vsun}) and (\ref{eq:omega}) into Eq.~(\ref{eq:bottlinger}), we obtain the system of conditional equations used in this work:

\begin{align}
\label{eq:bottlinger_final}
V_r ={}&
-\left(u_0\cos b\cos l
+v_0\cos b\sin l
+w_0\sin b\right)
\nonumber\\
&+R_0(R-R_0)\sin l\cos b\,\omega'_0
+\frac12 R_0(R-R_0)^2\sin l\cos b\,\omega''_0
\nonumber\\
&+rK\cos^2 b, \notag
\\[1ex]
V_l ={}&
v_0\cos l-u_0\sin l
+(R-R_0)(R_0\cos l-r\cos b)\omega'_0
\nonumber\\
&+\frac12(R-R_0)^2(R_0\cos l-r\cos b)\omega''_0
-r\omega_0\cos b,
\\[1ex]
V_b ={}&
w_0\cos b
-u_0\sin b\cos l
-v_0\sin b\sin l
\nonumber\\
&-R_0(R-R_0)\sin l\sin b\,\omega'_0
-\frac12 R_0(R-R_0)^2\sin l\sin b\,\omega''_0
\nonumber\\
&+rK\cos b\sin b. \notag
\end{align}

where the parameter $K$ represents a radial expansion or contraction component of the local stellar velocity field. It is introduced as an additional kinematic term motivated by the Ogorodnikov--Milne formalism and corresponds to the classical Oort $K$-term \citep{Ogorodnikov1965}. Although the Bottlinger model assumes circular differential rotation, the inclusion of $K$ allows us to test for possible systematic radial motions within the stellar population.

In the system of conditional equations used in this work, the Galactocentric distance $R$ is not treated as an independent parameter but is calculated for each star from Eq.~(\ref{eq:R}).

The unknown model parameters are the Solar peculiar velocity components $(u_0,v_0,w_0)$, the Solar Galactocentric distance $R_0$, the Galactic angular velocity parameters $\omega_0$, $\omega'_0$, and $\omega''_0$, and the radial expansion/contraction parameter $K$. The resulting system of conditional equations is nonlinear and therefore requires an iterative solution. For each star, Eq.~(\ref{eq:bottlinger_final}) provides three conditional equations corresponding to the observed velocity components $V_r$, $V_l$, and $V_b$.

\subsection{Generalised least-squares solution}
\label{sec:gls}

The model parameters are determined by simultaneously fitting all three velocity components to the observations using a nonlinear least-squares method. For convenience, the system of observation equations can be written in vector form as

\begin{align}
\label{eq:bottlinger_vector}
\mathbf{y}_i = \mathbf{A}_i \mathbf{p} + \boldsymbol{\varepsilon}_i,
\end{align}
where $\mathbf{y}_i$ is the vector of observed velocity components, $\mathbf{A}_i(\mathbf{p})$ is the design matrix constructed from Eq.~(\ref{eq:bottlinger_final}), $\mathbf{p}$ is the vector of unknown model parameters, and $\boldsymbol{\varepsilon}_i$ is the residual vector.

The observational residuals are affected by correlated uncertainties propagated from the Gaia astrometric solution. Therefore, the parameter vector $\mathbf{p}$ was estimated using generalized least squares (GLS), taking the full covariance matrix of the velocity components into account. Following \citet{GaiaCol2022, Akhmetov2026}, the covariance matrix for each star was decomposed using the Cholesky factorization, 
\begin{align}
\mathbf{C}_i=\mathbf{L}_i\mathbf{L}_i^{\mathrm T}, \qquad
\mathbf{K}_i=\mathbf{L}_i^{-1}, \notag
\end{align}
where $\mathbf{L}_i$ is the lower triangular Cholesky factor and $\mathbf{K}_i$ is the corresponding whitening matrix.

Premultiplying Eq.~(\ref{eq:bottlinger_vector}) by $\mathbf{K}_i$ yields the whitened system

\begin{align}
\mathbf{K}_i\mathbf{y}_i
=
\mathbf{K}_i\mathbf{A}_i\mathbf{p}
+
\mathbf{K}_i\boldsymbol{\varepsilon}_i, \notag
\end{align}
for which the transformed residuals have a unit covariance matrix. The parameter vector is therefore obtained by minimizing the sum of squared whitened residuals over all stars,

\begin{align}
\chi^2(\mathbf p)
=
\sum_i
\left\|
\mathbf K_i\mathbf y_i
-
\mathbf K_i\mathbf A_i\mathbf p
\right\|^2, \notag
\end{align}
where the norm is taken over the three velocity components of each star.

The resulting nonlinear optimization problem was solved using the Levenberg--Marquardt algorithm. All numerical computations were performed in C++ using the Eigen linear algebra library \citep*{eigenweb}.

\subsection{Spatial binning}
\label{sec:binning}

Although the GLS formalism described in Section~\ref{sec:gls} can be applied directly to individual stars, in this work we apply it to covariance-weighted averages computed within spatial bins. This approach suppresses small-scale fluctuations, improves the statistical stability of the solution, and reduces the computational cost while preserving the large-scale structure of the Galactic velocity field.

The sky was partitioned using the HEALPix (Hierarchical Equal Area iso-Latitude Pixelization) tessellation scheme \citep{Gorski2005} with $N_{\rm side}=16$, corresponding to a characteristic angular scale of approximately $3.66^\circ$. The heliocentric distance was divided into uniform bins with a width of $\Delta r=0.1$ kpc.

Within each spatial bin $Q$, the Galactic coordinates $(l, b, \varpi)$ and kinematic observables $(v_r,\mu_l,\mu_b)$ were combined into covariance-weighted averages using generalized least squares.

\begin{align}
\bar{\mathbf x}_Q
=
\mathbf C_Q
\sum_{i\in Q}
\mathbf C_i^{-1}\mathbf x_i, \notag
\end{align}
where the covariance matrix of the weighted mean is
\begin{align}
\mathbf C_Q
=
\left(
\sum_{i\in Q}
\mathbf C_i^{-1}
\right)^{-1}.  \notag
\end{align}

The heliocentric distance assigned to each spatial bin was then computed from the weighted mean parallax as $r=1/\bar{\varpi}$. Its uncertainty was propagated from the corresponding covariance matrix.

The covariance-weighted mean Galactic coordinates, proper motions, radial velocities, and their associated covariance matrices were subsequently used as input to the nonlinear GLS solution described in Section~\ref{sec:gls}.

The statistical uncertainties of the derived kinematic parameters were assessed using Monte Carlo simulations following the methodology developed and applied in our previous studies \citep{Akhmetov2026, Fedorov2026}. Multiple realizations of the stellar sample were generated by perturbing the measured parallaxes, proper motions, and radial velocities according to their full covariance matrices. The analysis procedure described above was then applied independently to each realization.

\section{Results}
\label{sec:res}

This section presents the fundamental Galactic parameters derived from the Bottlinger model, namely the solar Galactocentric distance $R_0$, the components of the Solar peculiar velocity $(u_0,v_0,w_0)$, the local angular velocity of Galactic rotation and its first two derivatives $(\omega_0,\omega'_0,\omega''_0)$, and the radial expansion/contraction parameter $K$.

The parameters were determined for a sequence of heliocentric distance limits, $r<R_s$, where $R_s$ denotes the maximum heliocentric distance of the stars included in the solution. This allows us to investigate the dependence of the derived parameters on the adopted sample volume and to identify the range of $R_s$ over which the solutions remain stable.

For each combination of the heliocentric distance limit $R_s$ and the vertical extent of the sample, the complete analysis described in Section~\ref{sec:met} was performed independently. The parameter values presented below were derived from the filtered Monte Carlo distributions, with the mean values used to characterise the solutions at each $R_s$ and the corresponding standard deviations used as Monte Carlo-based uncertainties. To reduce the influence of individual outlying realizations, a median absolute deviation (MAD)-based filter was applied independently to each parameter distribution. On average, this procedure rejected approximately $5-10\%$ of the Monte Carlo realizations.

The behaviour of the individual parameters as a function of the limiting heliocentric distance $R_s$ is discussed below.

\subsection{Galactocentric distance}

\begin{figure*}
\centering
\includegraphics[width=0.49\textwidth]{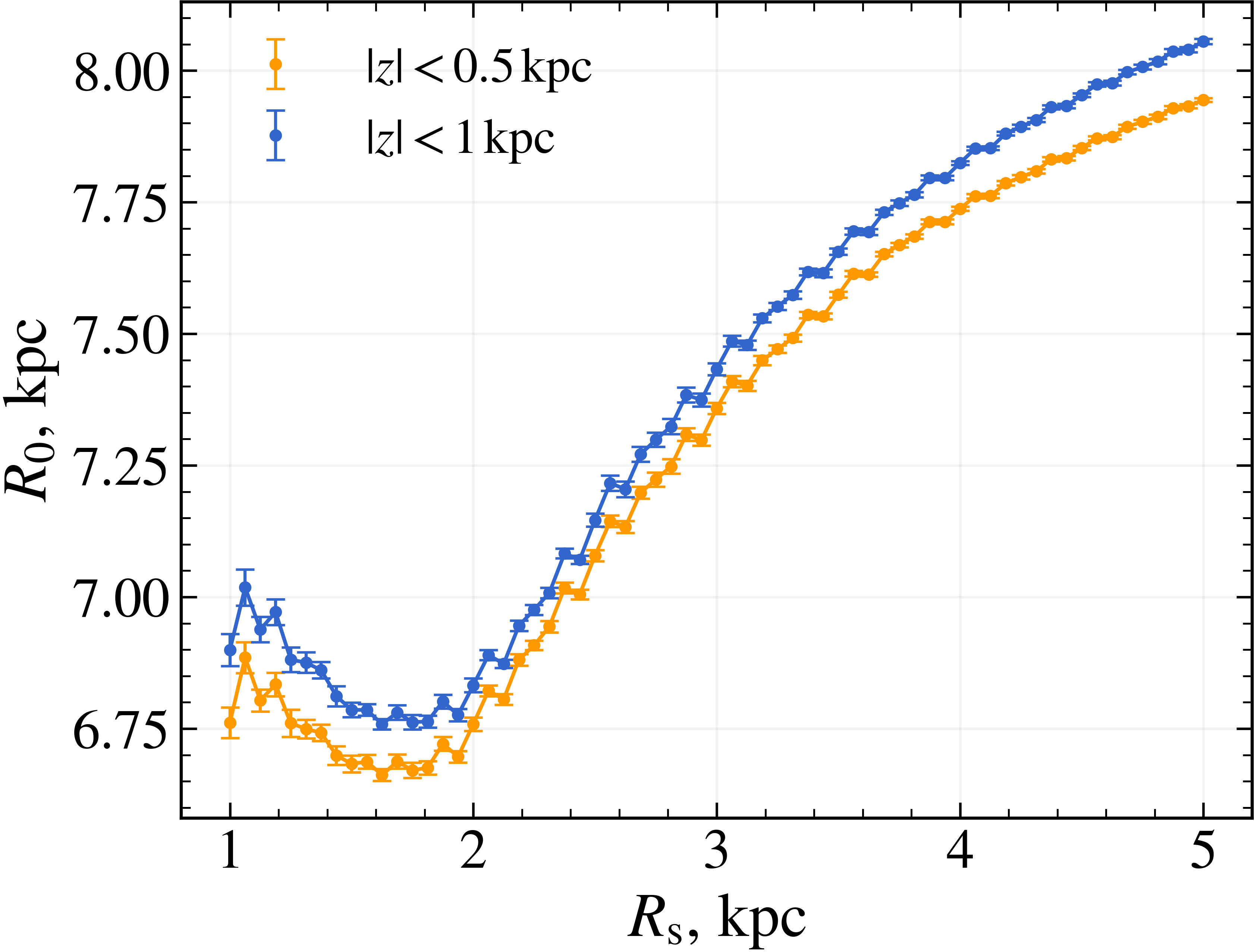}
\hfill
\includegraphics[width=0.49\textwidth]{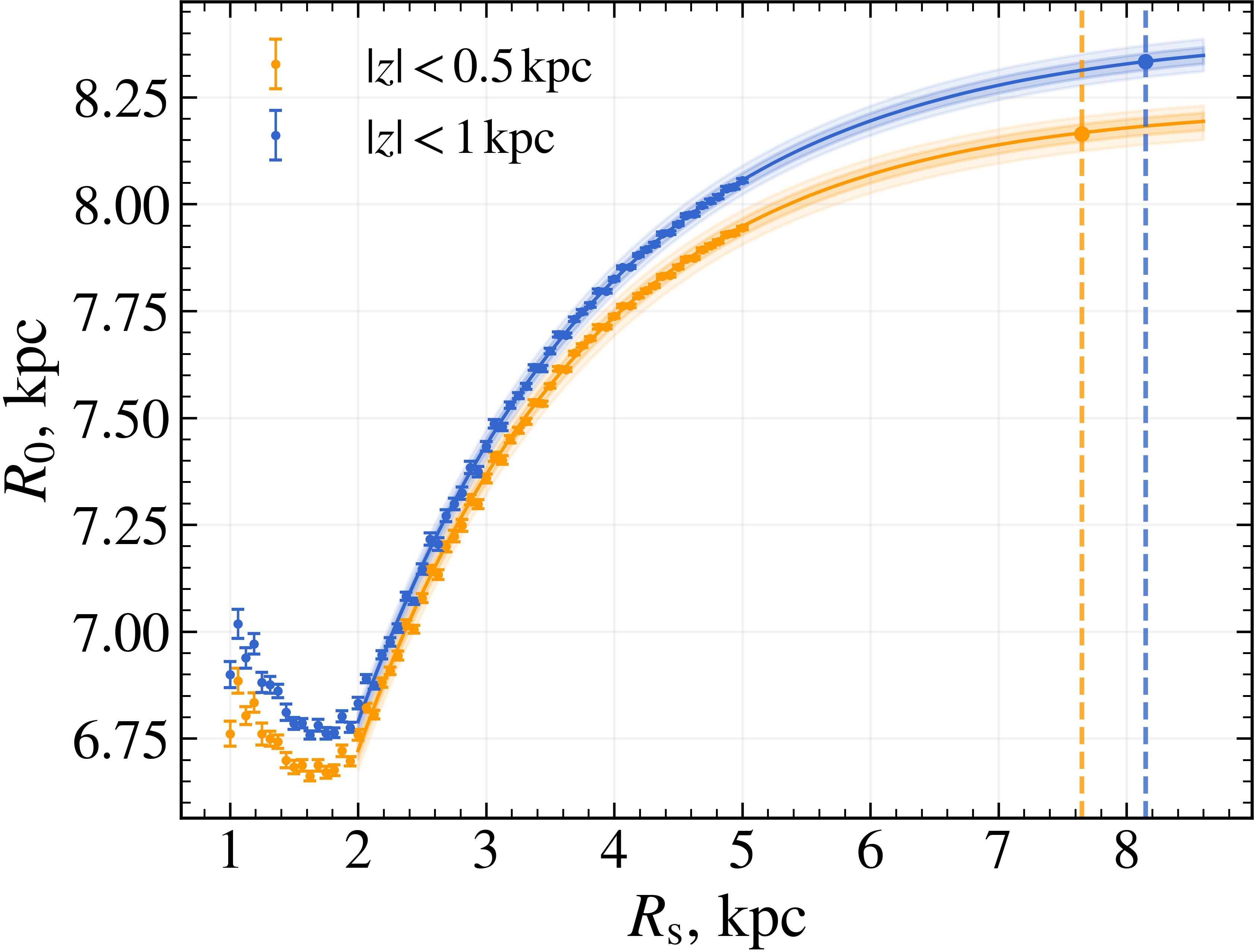}
\caption{\textit{Left}: Dependence of the derived solar Galactocentric distance $R_0$ on the heliocentric distance limit $R_s$ for the two vertical stellar subsamples. \textit{Right}: Exponential extrapolation of the $R_0(R_s)$ relation beyond the distance range directly sampled by Gaia DR3. The shaded regions indicate the uncertainty bands of the fitted relation corresponding to the 68\% and 95\% confidence levels, while the vertical dashed lines indicate the adopted stabilization distances used to estimate $R_0$.}
\label{fig:R0}
\end{figure*}

Fig.~\ref{fig:R0} (\textit{left panel}) shows the dependence of the derived solar Galactocentric distance $R_0$ on the adopted heliocentric distance limit of the stellar samples, $R_s$. The estimated values gradually approach an asymptotic regime as $R_s$ increases, although convergence is not fully reached within the distance range directly sampled by the Gaia DR3 data.

Tests performed on Gaia-like mock catalogues (Fig.~\ref{fig:R0_mock}) demonstrate that, when sufficiently large heliocentric distances are included, the recovered values of $R_0$ converge towards a stable value. However, the Gaia DR3 sample, extending the analysis to similarly large distances is not justified because the astrometric precision of Gaia decreases with increasing distance, leading to increasingly large relative uncertainties in the measured parallaxes. Consequently distances obtained by direct inversion of Gaia parallaxes become increasingly unreliable at large heliocentric distances \citep{BailerJones2023}. Therefore, the asymptotic behaviour of the $R_0(R_s)$ relation was estimated by fitting and extrapolating the observed trend (Fig.~\ref{fig:R0}, \textit{right panel}).

The dependence was approximated by
\begin{equation}
R_0(R_s)=R_{\infty}+A
\exp\left(-\frac{R_s}{r_0}\right),
\notag
\end{equation}
where $R_{\infty}$, $A$, and $r_0$ are free parameters determined by fitting the exponential model to the derived $R_0(R_s)$ relation. The fitted function was then extrapolated beyond the observed distance range to estimate the asymptotic behaviour of the solution.

The value of $R_0$ was evaluated from the fitted relation at the onset of the stabilization region, defined as the smallest $R_s$ for which the fitted relation approaches the asymptotic value $R_{\infty}$ within the residual scatter of the fit, estimated using the median absolute deviation (MAD). The statistical uncertainty was estimated using Monte Carlo realizations, in which the derived $R_0(R_s)$ values were perturbed according to their individual uncertainties and the fitting and stabilization procedure was repeated; the standard deviation of the resulting distribution of extrapolated $R_0$ values was adopted as $\sigma_{\mathrm{MC}}$. The robustness of the extrapolation was assessed by varying the MAD threshold, stabilization-window size, and minimum heliocentric distance included in the fit, with the resulting scatter adopted as the methodological contribution, $\sigma_{\mathrm{sens}}$. The total uncertainty was then obtained by combining these contributions in quadrature, $\sigma_{R_0}=\sqrt{\sigma_{\mathrm{MC}}^2+\sigma_{\mathrm{sens}}^2}$, and the uncertainty bands shown in Fig.~\ref{fig:R0} (\textit{right}) include both contributions.

Applying this procedure yields $R_0 = 8.165 \pm 0.024$~kpc for the sample with $|z|<0.5$~kpc and $R_0 = 8.333 \pm 0.022$~kpc for the sample with $|z|<1.0$~kpc.

\subsection{Solar peculiar motion}
\label{sec:solar_motion}

\begin{figure*}
\centering
\includegraphics[width=0.33\textwidth]{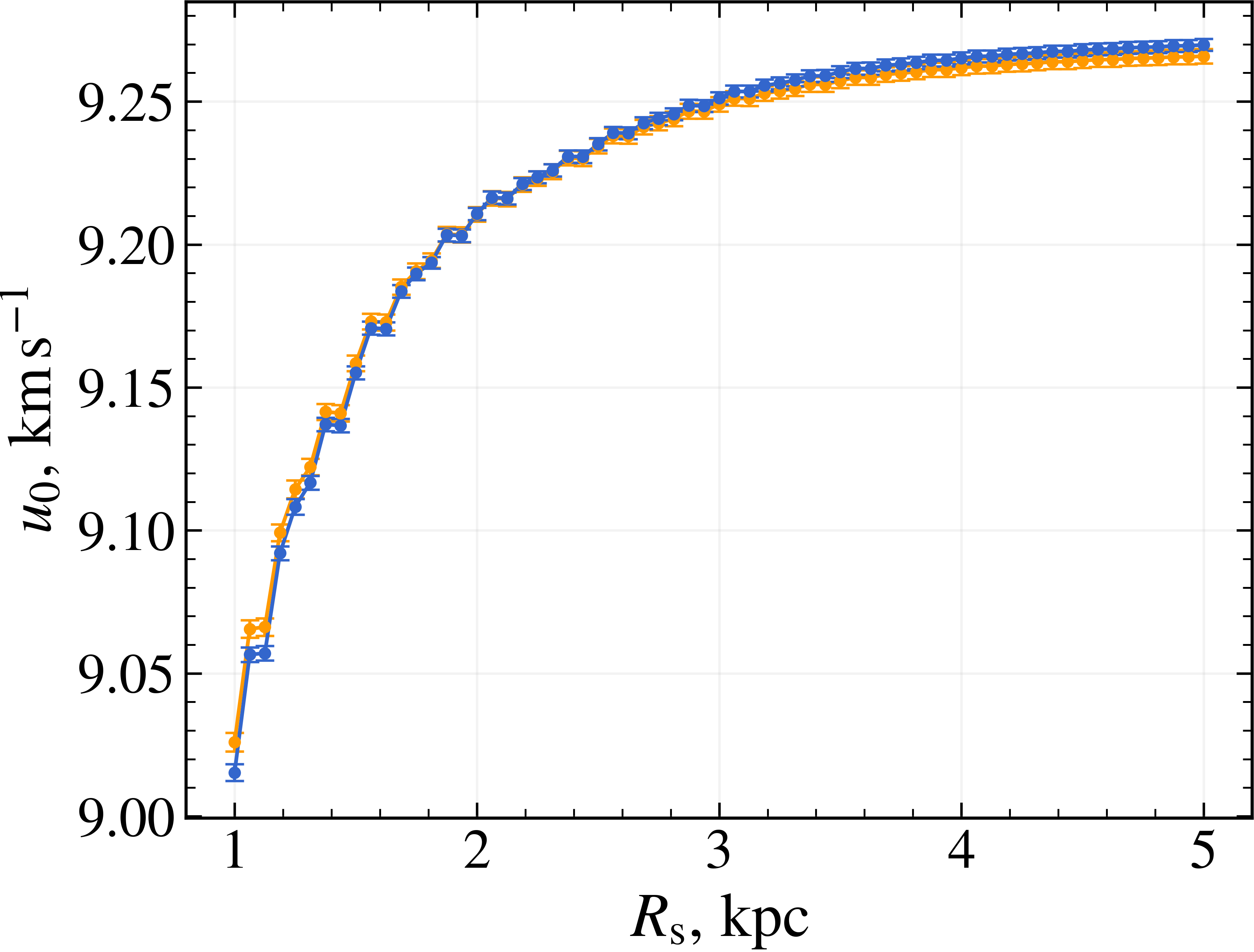}
\hfill
\includegraphics[width=0.33\textwidth]{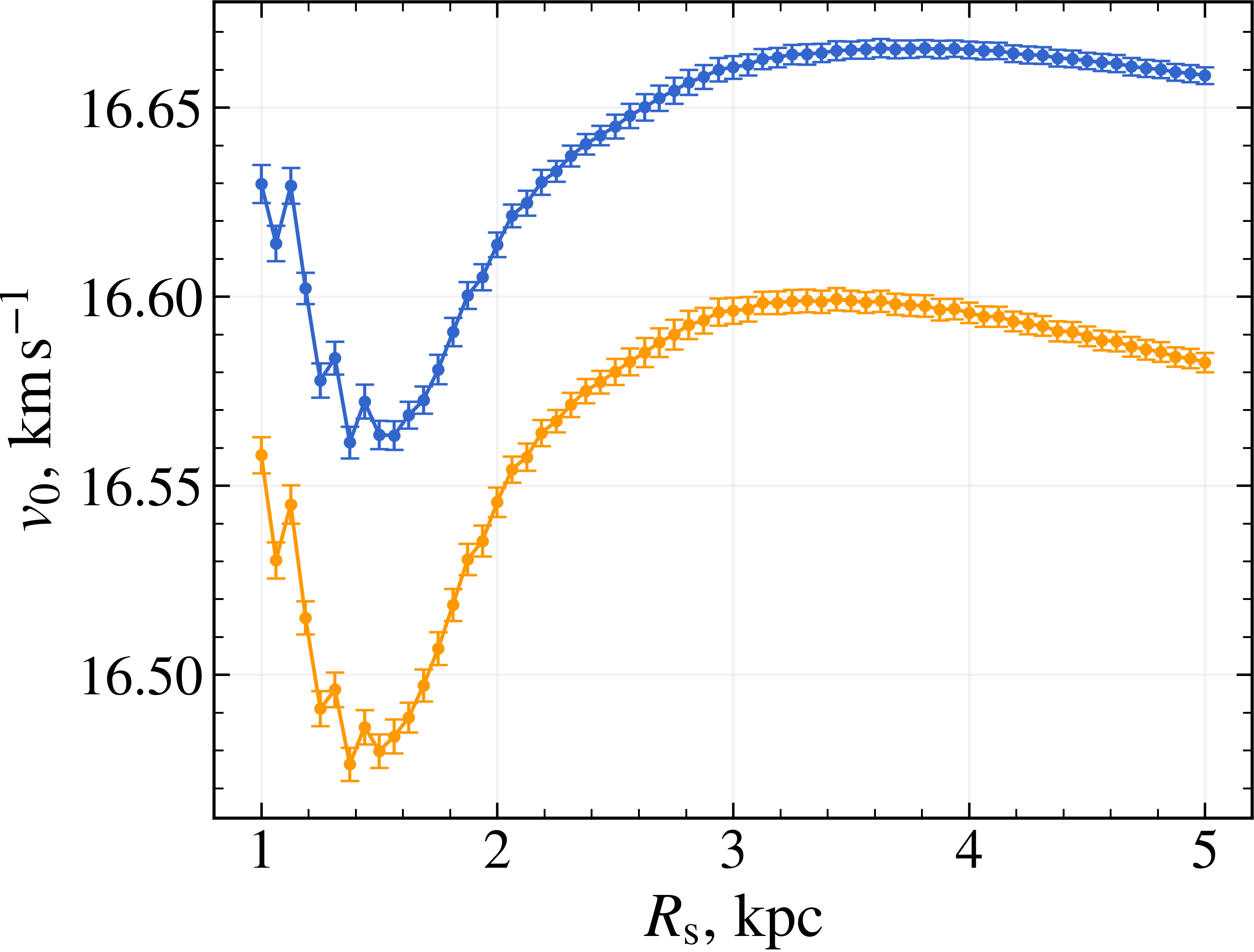}
\hfill
\includegraphics[width=0.33\textwidth]{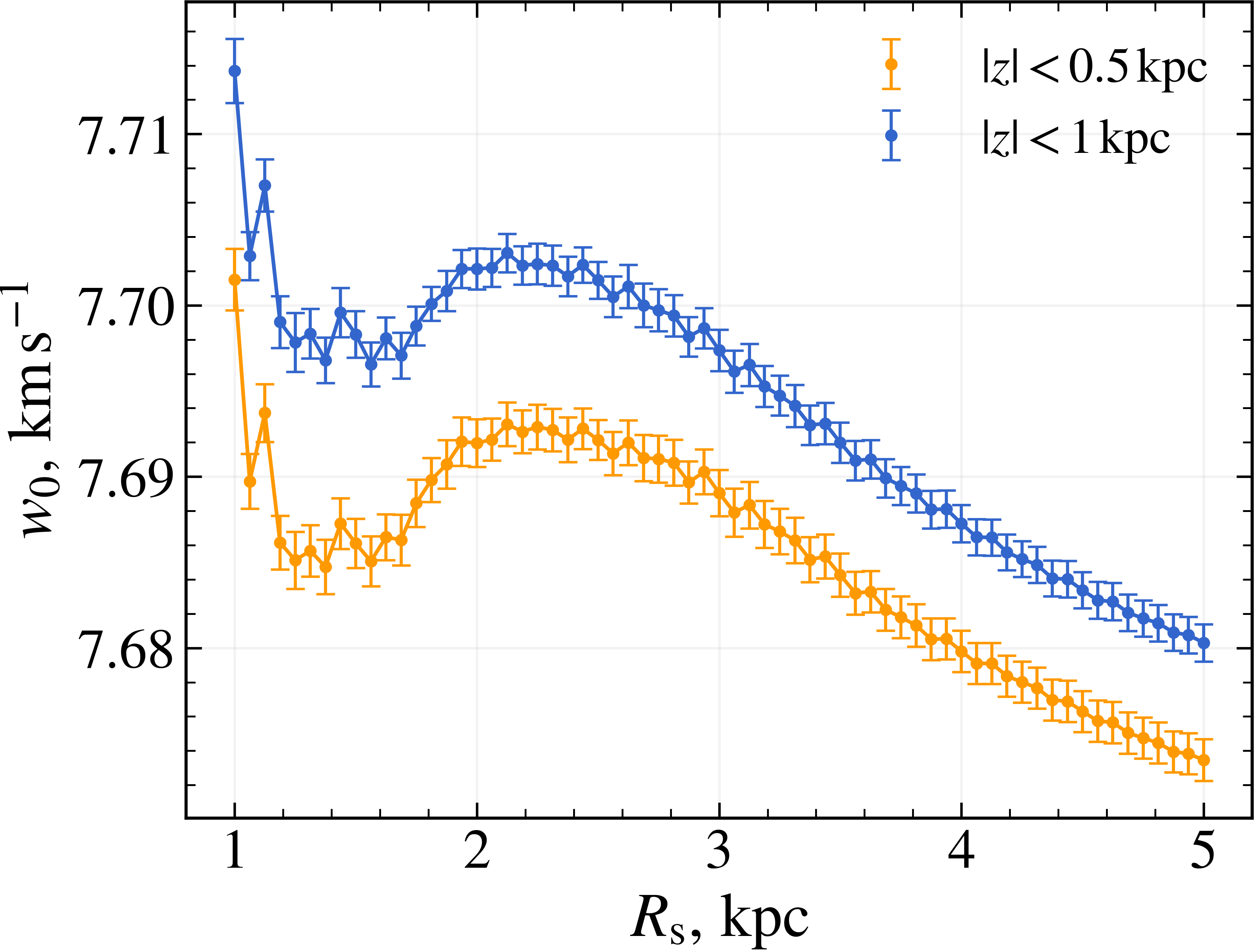}
\caption{Dependence of the Solar peculiar velocity components on the adopted heliocentric distance limit $R_s$. From left to right: the radial ($u_0$), tangential ($v_0$), and vertical ($w_0$) components for the two stellar subsamples defined by their vertical extent.}
\label{fig:solar_motion}
\end{figure*}

Fig.~\ref{fig:solar_motion} shows the dependence of the Solar peculiar velocity components on the adopted heliocentric distance limit $R_s$. In contrast to $R_0$, all three components become nearly constant once the sample extends beyond approximately 2--3~kpc, indicating that they are considerably less sensitive to the adopted distance limit.

The radial component $u_0$ exhibits the largest systematic variation, increasing by about $0.25~\mathrm{km\,s^{-1}}$ over the analysed range of $R_s$. The tangential component $v_0$ varies by less than $0.1~\mathrm{km\,s^{-1}}$, while the vertical component $w_0$ remains nearly constant throughout the investigated range.

The final parameter estimates were obtained from the plateau region of each dependence, retaining only measurements beyond the adopted stabilization radius. Outlying solutions were removed using an iterative $3\sigma$ clipping procedure based on the median absolute deviation (MAD), and the final value was taken as the median of the remaining measurements. The uncertainty was estimated by combining the robust scatter of the plateau values, $\sigma_{\mathrm{MAD}}$, obtained from the MAD and converted to the equivalent Gaussian standard deviation, with the characteristic Monte Carlo uncertainty of the retained measurements, $\sigma_{\mathrm{MC}}$, giving $\sigma=\sqrt{\sigma_{\mathrm{MAD}}^{2}+\sigma_{\mathrm{MC}}^{2}}$.

The Solar peculiar motion derived for the $|z|<0.5$~kpc subsample is $(u_0,v_0,w_0)=(9.261\pm0.006,\ 16.593\pm0.008,\ 7.686\pm0.008)~\mathrm{km\,s^{-1}}$. Increasing the vertical extent of the sample to $|z|<1.0$~kpc results in changes of less than $0.1~\mathrm{km\,s^{-1}}$ in each component. This small variation demonstrates the stability of the inferred Solar motion against the adopted vertical cut. The resulting parameters are listed in Table~\ref{tab:results}.

\subsection{Local angular velocity and its radial derivatives}
\label{sec:omega}

\begin{figure*}
\centering
\includegraphics[width=0.33\textwidth]{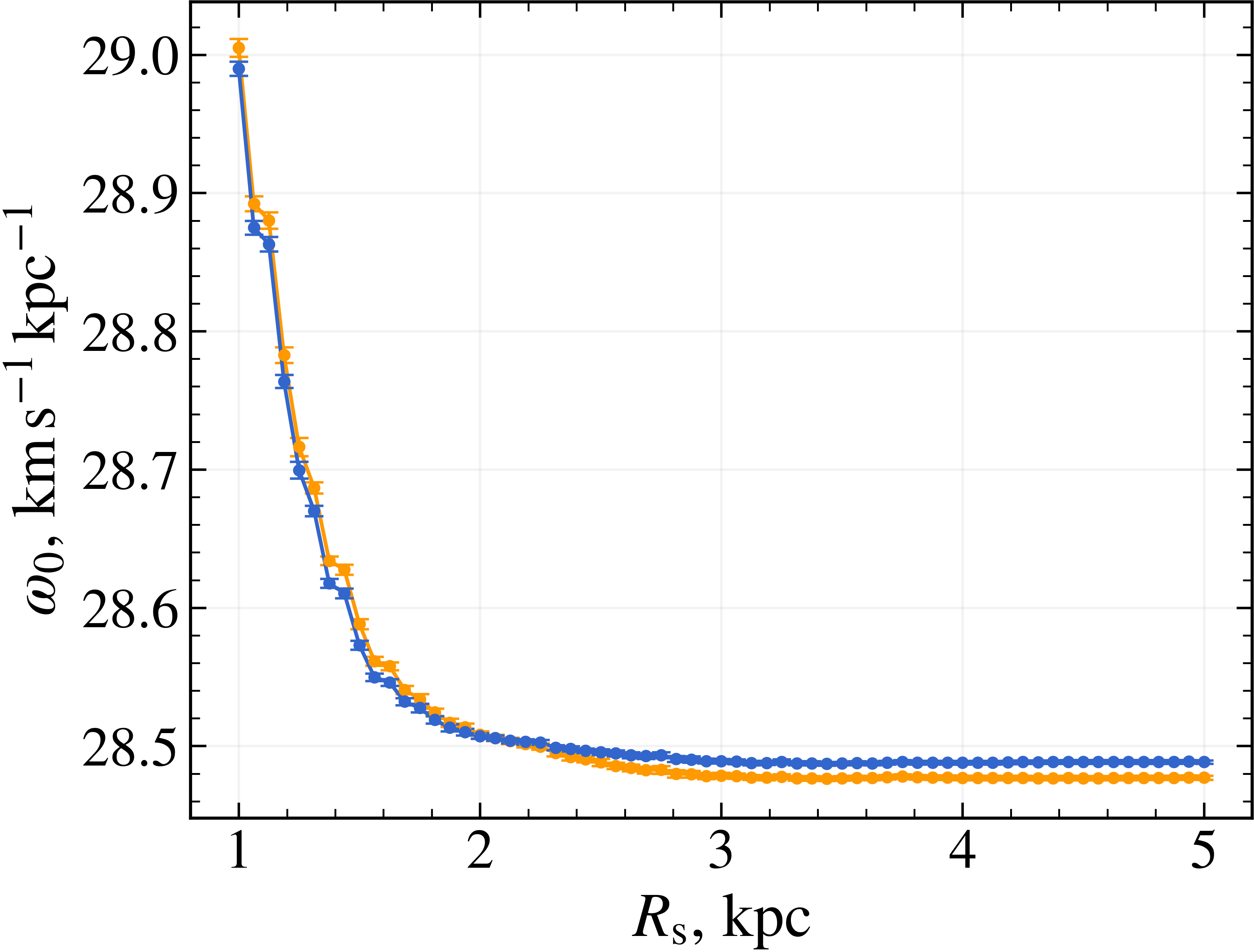}
\hfill
\includegraphics[width=0.33\textwidth]{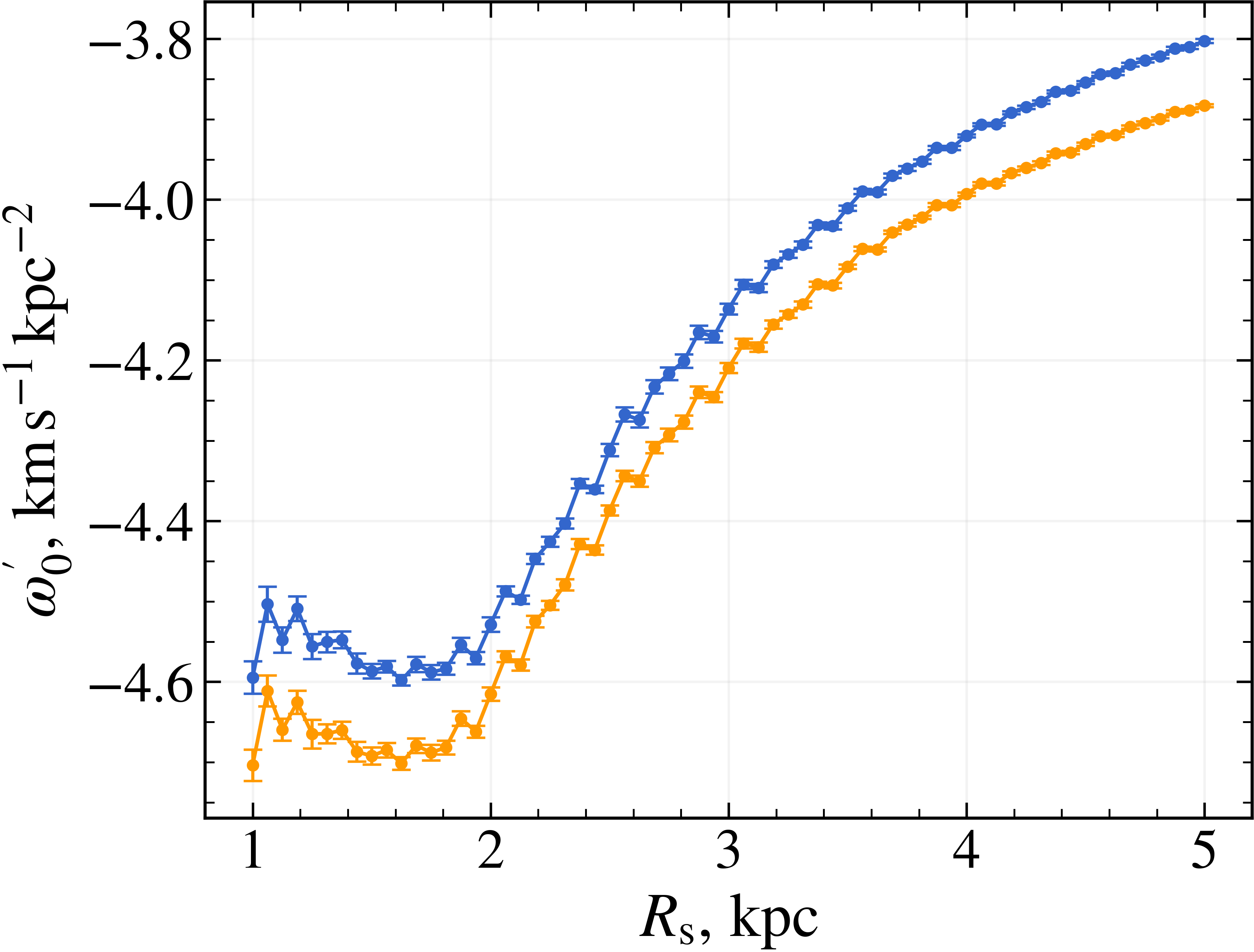}
\hfill
\includegraphics[width=0.33\textwidth]{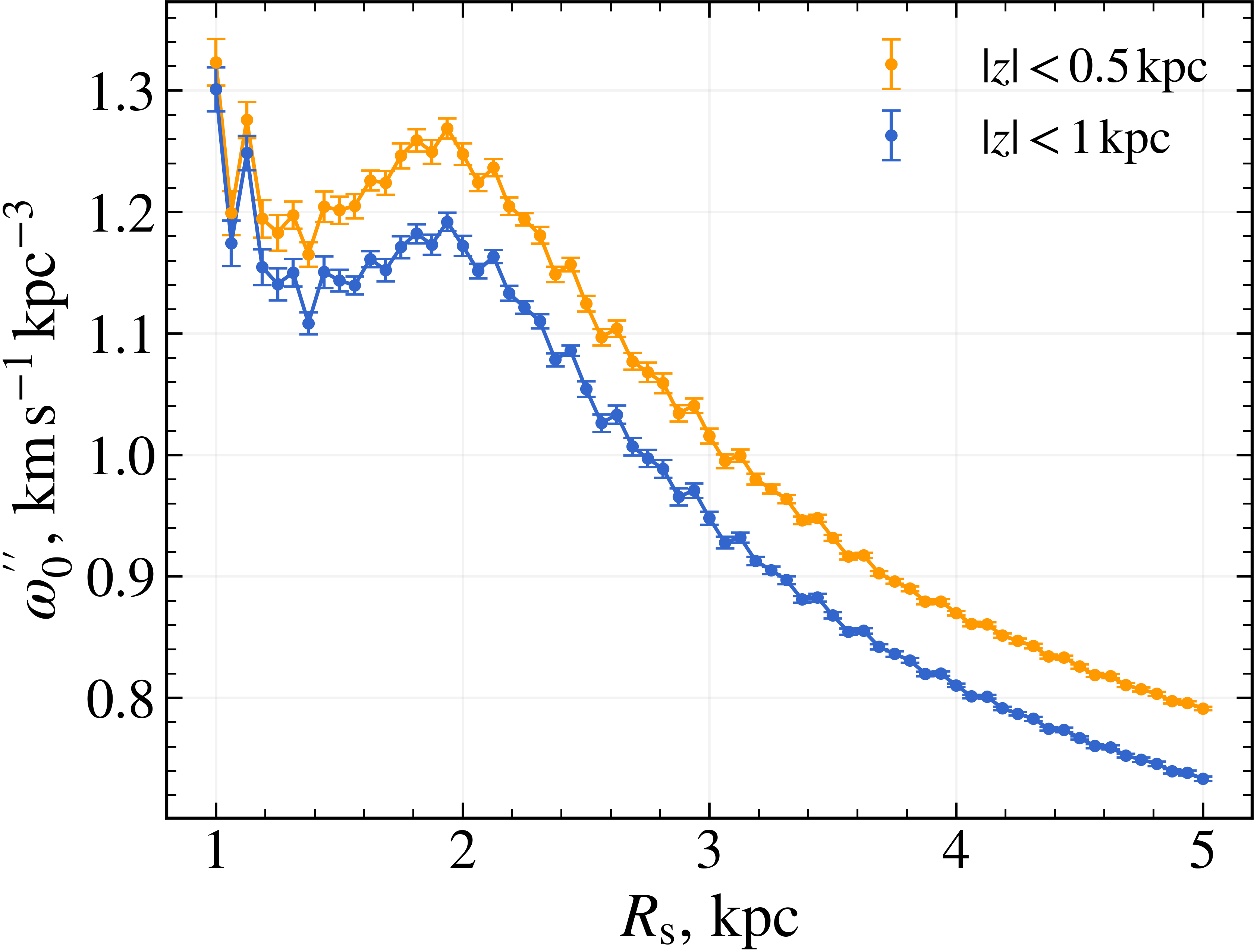}
\caption{Dependence of the local angular velocity of Galactic rotation, $\omega_0$ (left), its first radial derivative, $\omega'_0$ (centre), and second radial derivative, $\omega''_0$ (right), on the heliocentric distance limit $R_s$ for the two vertical stellar subsamples.}
\label{fig:rotation}
\end{figure*}

Fig.~\ref{fig:rotation} presents the dependence of the Galactic angular velocity $\omega_0$ and its first two radial derivatives, $\omega'_0$ and $\omega''_0$, on the adopted heliocentric distance limit $R_s$. The behaviour of these parameters differs significantly from that of the Solar peculiar velocity components, reflecting their sensitivity to the spatial scale over which the Galactic rotation field is constrained.

The angular velocity $\omega_0$ shows rapid convergence with increasing $R_s$ and becomes nearly independent of the adopted sample volume for $R_s \gtrsim 2$--$3$~kpc. The solutions obtained for the two vertical subsamples remain mutually consistent over the entire analysed range, indicating that the determination of the local angular velocity is robust against variations in the vertical selection of stars.

In contrast, the radial derivatives $\omega'_0$ and $\omega''_0$ show a stronger dependence on the adopted spatial extent of the sample. Both parameters continue to vary with increasing $R_s$ and do not reach a clearly defined asymptotic regime within the analysed distance range. This behaviour indicates that the derivatives of the angular velocity are more sensitive to the radial coverage of the sample than the local value of $\omega_0$ itself.

This difference is expected because $\omega_0$ describes the angular velocity at the Solar Galactocentric distance, whereas $\omega'_0$ and $\omega''_0$ characterize the local slope and curvature of the angular-velocity profile. Consequently, their determination depends on the radial range over which the Taylor expansion in Eq.~(\ref{eq:omega}) is constrained and remains sensitive to the adopted spatial volume.

The final value of $\omega_0$ was determined from the plateau region of the $\omega_0(R_s)$ dependence using the same plateau-selection and uncertainty-estimation procedure described in the previous subsection. Since neither $\omega'_0$ nor $\omega''_0$ reaches a comparable plateau, no analogous stabilization analysis was applied to these parameters.

The resulting local angular velocity for the sample with $|z|<0.5$~kpc is
$\omega_0 = 28.477 \pm 0.002~\mathrm{km\,s^{-1}\,kpc^{-1}}$.
The corresponding value obtained for the sample with $|z|<1.0$~kpc is consistent within the estimated uncertainties, indicating that the determination of $\omega_0$ is robust with respect to the adopted vertical selection. The adopted values of the Galactic rotation parameters are summarized in Table~\ref{tab:results}.

\subsection{Radial expansion/contraction parameter}

\begin{figure}
\centering
\includegraphics[width=0.9\columnwidth]{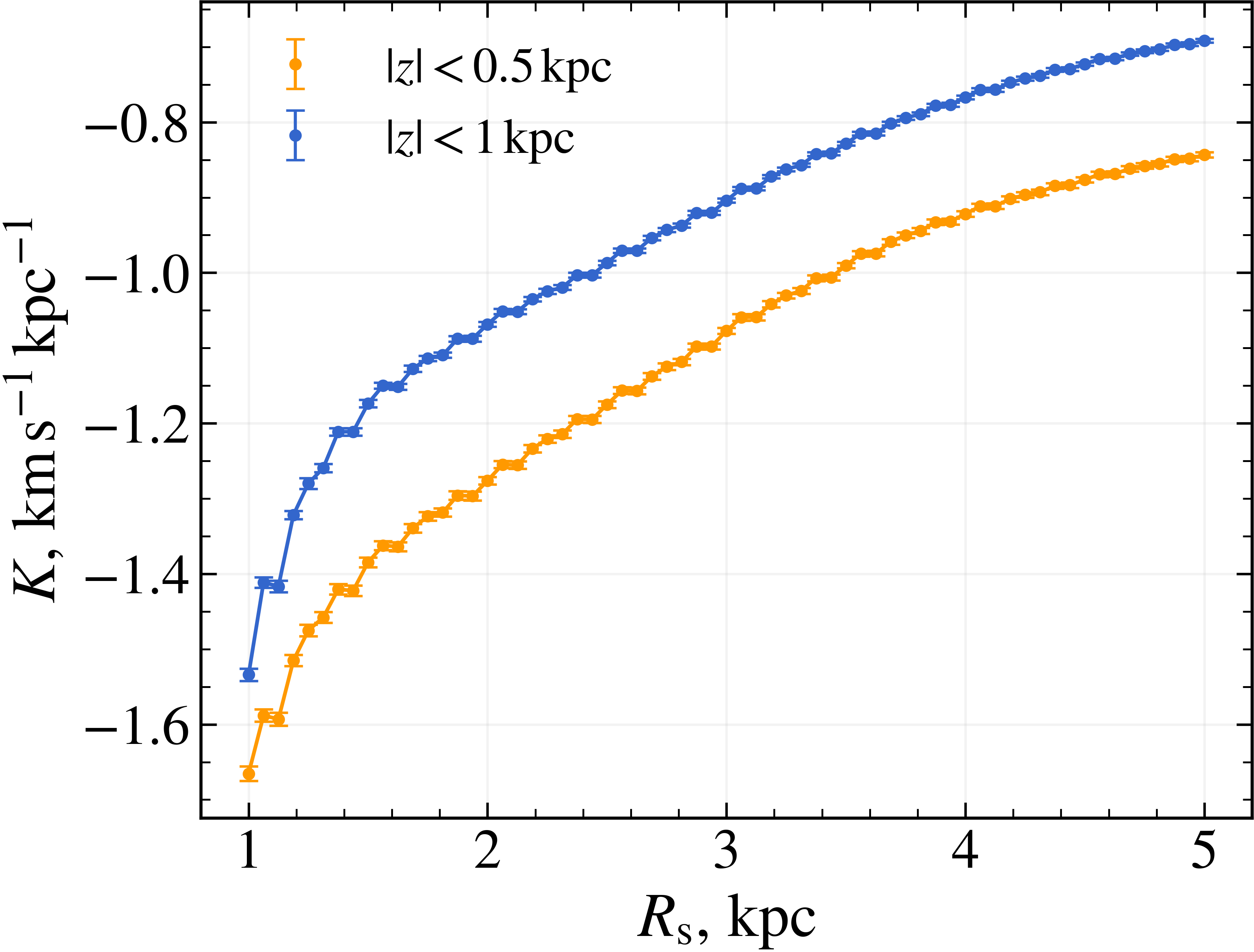}
\caption{Dependence of the radial expansion/contraction parameter $K$ on the heliocentric distance limit $R_s$ for the two vertical subsamples.}
\label{fig:K}
\end{figure}

Fig.~\ref{fig:K} shows the dependence of the radial expansion parameter $K$ on the heliocentric distance limit $R_s$. In contrast to the local angular velocity $\omega_0$, the parameter $K$ exhibits a pronounced dependence on the adopted spatial extent of the sample, indicating that its determination is sensitive to the radial coverage and spatial distribution of the tracer population.

From a physical point of view, $K=0$ corresponds to the absence of a systematic radial expansion or contraction of the stellar velocity field. Thus, a value consistent with zero indicates that no net radial deformation of the velocity field is detected within the adopted model. Significant deviations from zero, on the other hand, indicate the presence of systematic  stellar radial motions/migration and may point to departures from a stationary radial velocity field.

At small values of $R_s$, the derived values of $K$ show substantial negative deviations from zero, reaching approximately $-1.6$-$-1.4~\mathrm{km\,s^{-1}\,kpc^{-1}}$. Such values indicate a significant radial contraction of the velocity field within the inner part of the sampled volume, rather than a configuration consistent with $K=0$. As the heliocentric distance limit increases, $K$ increases monotonically, that is, its magnitude decreases and the inferred radial contraction becomes progressively weaker. This behaviour is particularly pronounced at $R_s\lesssim2$--$3$~kpc.

The two vertical subsamples exhibit very similar trends but remain systematically offset from each other over the entire investigated range. The sample with $|z|<1.0$~kpc yields consistently larger values of $K$ than the thin-disk-dominated sample with $|z|<0.5$~kpc. At the largest distances considered, $K$ approaches approximately $-0.8$--$-0.85~\mathrm{km\,s^{-1}\,kpc^{-1}}$ for the $|z|<0.5$~kpc sample and $-0.65$--$-0.7~\mathrm{km\,s^{-1}\,kpc^{-1}}$ for the $|z|<1.0$~kpc sample.

Although the magnitude of $K$ decreases substantially with increasing $R_s$, the parameter remains significantly different from zero even at the largest distances considered. Thus, the radial velocity field is not fully consistent with a stationary configuration over the spatial scales probed by the sample. The convergence towards less negative values suggests that the strong radial contraction inferred at small spatial scales is progressively diluted when stars at larger heliocentric distances are included. Consequently, a unique stabilization value of $K$ cannot be determined in the same manner as for the Solar peculiar motion and the local angular velocity. The values of $K$ should therefore be interpreted as scale-dependent characteristics of the radial velocity field. The systematic offset between the two vertical subsamples further suggests that the inferred radial motions depend on the vertical composition of the tracer population.

\section{Discussion}
\label{sec:dis}

The dependence of the fitted parameters on the adopted heliocentric distance limit reveals different responses to the spatial extent and vertical selection of the stellar sample. While some parameters approach stable values, others retain a systematic dependence on the adopted sample volume, indicating that the resulting Galactic parameters are sensitive to both the spatial coverage and the properties of the stellar tracers.

\subsection{Comparison with previous studies}

\begin{figure}
\centering
\includegraphics[width=0.9\columnwidth]{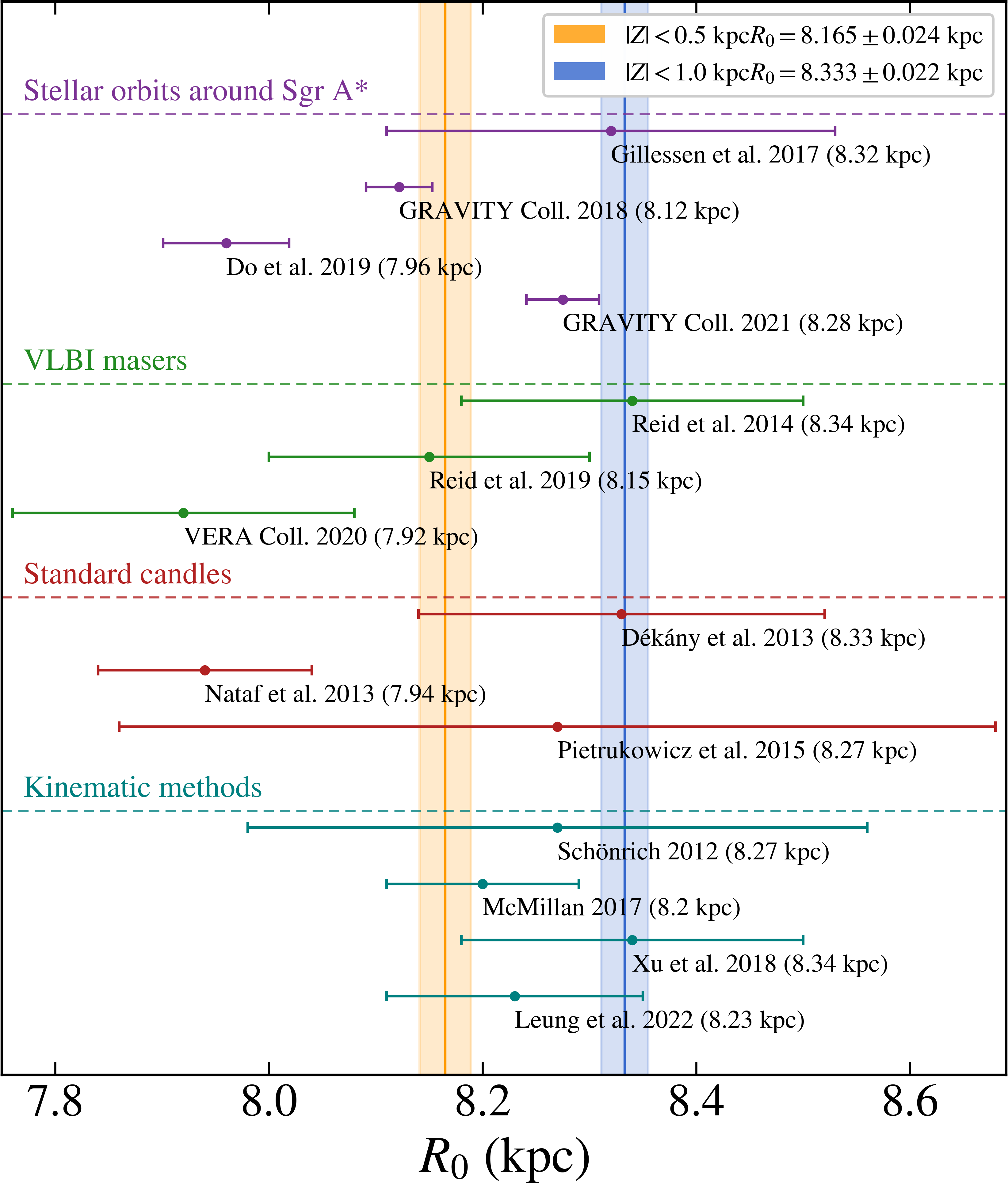}
\caption{A selection of historical measurements of $R_0$ obtained using different observational methods \citep{Gillessen2017,Do2019,Gravity2018,Gravity2021,Reid2014,Reid2019,VERA2020,Dekany2013,Nataf2013,Pietrukowicz2015,McMillan2017,Schonrich2012,Xu2018,Leung2022}. The vertical orange and blue lines and the corresponding shaded bands indicate the results and their uncertainties obtained in this work.}
\label{fig:R0_comp}
\end{figure}

\begin{table*}
\caption{Comparison of the Galactic kinematic parameters derived in this work with selected recent determinations from the literature.}
\label{tab:comparison}
\centering
\begin{tabular}{lccccc}
\hline
Study & $R_0$ (kpc) & $u_0$ (km\,s$^{-1}$) & $v_0$ (km\,s$^{-1}$) & $w_0$ (km\,s$^{-1}$) & $\omega_0$ (km\,s$^{-1}$\,kpc$^{-1}$)\\
\hline

Reid et al. (2019) & $8.15\pm0.15$ & $10.6\pm1.2$ & $10.7\pm6.0$ & $7.6\pm0.7$ & $30.32\pm0.27$ \\

McMillan (2017) & $8.20\pm0.09$ & $8.6\pm0.9$ & $13.9\pm1.0$  & $7.1\pm1.0$  & $\sim26.7$ \\

Schönrich et al. (2010,2012) & $8.27\pm0.29$ & $11.1\pm0.69$  & $12.24\pm0.47$  & $7.25\pm0.37$  & $\sim28.8$  \\

Xu et al. (2018) & $8.35\pm0.18$ & $13.3\pm2.6$ & $17.0\pm8.0$  & $8.6\pm0.6$  & $\sim28.7$  \\

Fedorov et al. (2021) & $8.0$ (fixed) & $10.53\pm0.04$  & $25.28\pm0.04$  & $7.84\pm0.04$  & $\sim28.38$  \\

Bobylev et al. (2023) & $8.1$ (fixed) & $9.39\pm0.40$ & $15.96\pm0.55$ & $6.88\pm0.30$  & $27.87\pm0.09$  \\

\textbf{This work (|z| < 0.5 kpc)} & $8.165\pm0.024$ & $9.261\pm0.006$ & $16.593\pm0.008$ & $7.686\pm0.008$ & $28.477\pm0.002$ \\

\textbf{This work (|z| < 1.0 kpc)} & $8.333\pm0.022$ & $9.264\pm0.006$ & $16.663\pm0.004$ & $7.697\pm0.008$ & $28.488\pm0.001$ \\

\hline
\end{tabular}

\parbox{\textwidth}{%
\vspace{0.6em}
\footnotesize
\textbf{Notes.}The literature values are taken from
\citet{Reid2019},
\citet{McMillan2017},
\citet{Schonrich2010},
\citet{Schonrich2012},
\citet{Xu2018},
\citet{Fedorov2021},
and \citet{Bobylev2023}.
For studies in which $R_0$ was fixed during the fitting procedure, this is indicated in the second column. Values of $\omega_0$ not explicitly reported in the original publications were derived from the published rotation parameters.
}
\end{table*}

The recovered Solar Galactocentric distance is in good agreement with recent determinations obtained using independent observational techniques. In particular, our results are consistent with the GRAVITY Collaboration estimate of $R_0 = 8.275 \pm 0.034$~kpc \citep{Gravity2021}, the VLBI maser determination of $R_0 = 8.15 \pm 0.15$~kpc by \citet{Reid2019}, the dynamical estimate of $R_0 = 8.20 \pm 0.09$~kpc obtained by \citet{McMillan2017}, and the Galactic-bar determination of $R_0 = 8.23 \pm 0.12$~kpc reported by \citet{Leung2022}. Thus, both of our solutions lie within the range of values established by different observational methods and are broadly consistent with recent estimates, which generally place $R_0$ at approximately $8.1$--$8.3$~kpc \citep{BlandHawthorn2016,McMillan2017,Reid2019}. At the same time, the systematic increase in the estimated $R_0$ from the $|z|<0.5$~kpc to the $|z|<1.0$~kpc sample, corresponding to a difference of $\Delta R_0 \simeq 0.17$~kpc, indicates that the inferred Galactocentric distance depends to some extent on the adopted tracer population. This difference is comparable to the dispersion among many recent kinematic determinations summarized in Fig.~\ref{fig:R0_comp} and Table~\ref{tab:comparison}. Extending the vertical range changes the relative contribution of stellar populations with different kinematic properties, and similar tracer-dependent variations in Galactic parameters have been reported previously \citep{DeGrijs2017}. These differences may therefore contribute to the dispersion among published values of $R_0$, in addition to differences in the adopted sample selection and underlying modelling techniques.

The derived components of the Solar peculiar velocity are generally consistent with previous determinations summarized in Table~\ref{tab:comparison}, although noticeable differences are found for the tangential component. The recovered values of $u_0 = 9.26$--$9.27$~km\,s$^{-1}$ and $w_0 = 7.69$~km\,s$^{-1}$ are in close agreement with the estimates of \citet{McMillan2017} ($u_0=8.6\pm0.9$, $w_0=7.1\pm1.0$~km\,s$^{-1}$) and \citet{Bobylev2023} ($u_0=9.39\pm0.40$, $w_0=6.88\pm0.30$~km\,s$^{-1}$). The recovered tangential component, $v_0 = 16.59$--$16.66$~km\,s$^{-1}$, is higher than the classical value of $12.24\pm0.47$~km\,s$^{-1}$ reported by \citet{Schonrich2010}, but is consistent with the more recent determinations of $17.0\pm8.0$~km\,s$^{-1}$ by \citet{Xu2018} and $15.96\pm0.55$~km\,s$^{-1}$ by \citet{Bobylev2023}. Such differences are not unexpected, since the tangential component is particularly sensitive to the adopted Galactic rotation model and to the assumed circular velocity of the Local Standard of Rest \citep{Schonrich2012}. More generally, the Solar peculiar velocity is a local kinematic quantity, but its inferred components may still depend on the adopted tracer population and reference frame \citep{Katz2023}.

The local angular velocity of Galactic rotation derived in this work, $\omega_0 = 28.48$--$28.49$~km\,s$^{-1}$\,kpc$^{-1}$, is in good agreement with recent kinematic determinations summarized in Table~\ref{tab:comparison}. In particular, it is consistent with the values reported by \citet{Xu2018} ($\omega_0 \approx 28.7$~km\,s$^{-1}$\,kpc$^{-1}$), \citet{Fedorov2021} ($\omega_0 \approx 28.38$~km\,s$^{-1}$\,kpc$^{-1}$), and \citet{Bobylev2023} ($27.87\pm0.09$~km\,s$^{-1}$\,kpc$^{-1}$). The somewhat higher value obtained by \citet{Reid2019} ($30.32\pm0.27$~km\,s$^{-1}$\,kpc$^{-1}$) may reflect differences in the adopted tracer population and modelling approach. Overall, the close agreement with previous studies indicates that $\omega_0$ is among the most robust parameters recovered within the Bottlinger formalism.

In contrast, the radial derivatives $\omega'_0$ and $\omega''_0$, as well as the radial expansion parameter $K$, show a stronger dependence on the spatial extent of the sample. Recent Gaia DR3 analyses have demonstrated that velocity gradients and streaming motions vary systematically across the Galactic disc, reflecting the presence of non-axisymmetric structures and spatially varying kinematics rather than a single globally smooth velocity field \citep{Fedorov2023,Akhmetov2024}. The observed variation of $\omega'_0$ and $\omega''_0$ with the adopted heliocentric distance limit is therefore consistent with the fact that these parameters describe spatial gradients of the rotation field. Similarly, the lack of convergence of $K$ within the analysed distance range and its systematic variation with increasing sample volume are consistent with this spatially varying kinematic structure, including regions of local compression and expansion associated with spiral arms and the Galactic bar \citep{Denyshchenko2024}. These results suggest that the parameters describing velocity gradients and radial streaming are sensitive to the spatial scale of the analysed sample and should not necessarily be interpreted as global characteristics of the Galactic disc.

\subsection{Application to Gaia-like mock catalogues}
\label{sec:validation}

\begin{figure*}
\centering
\includegraphics[width=0.49\textwidth]{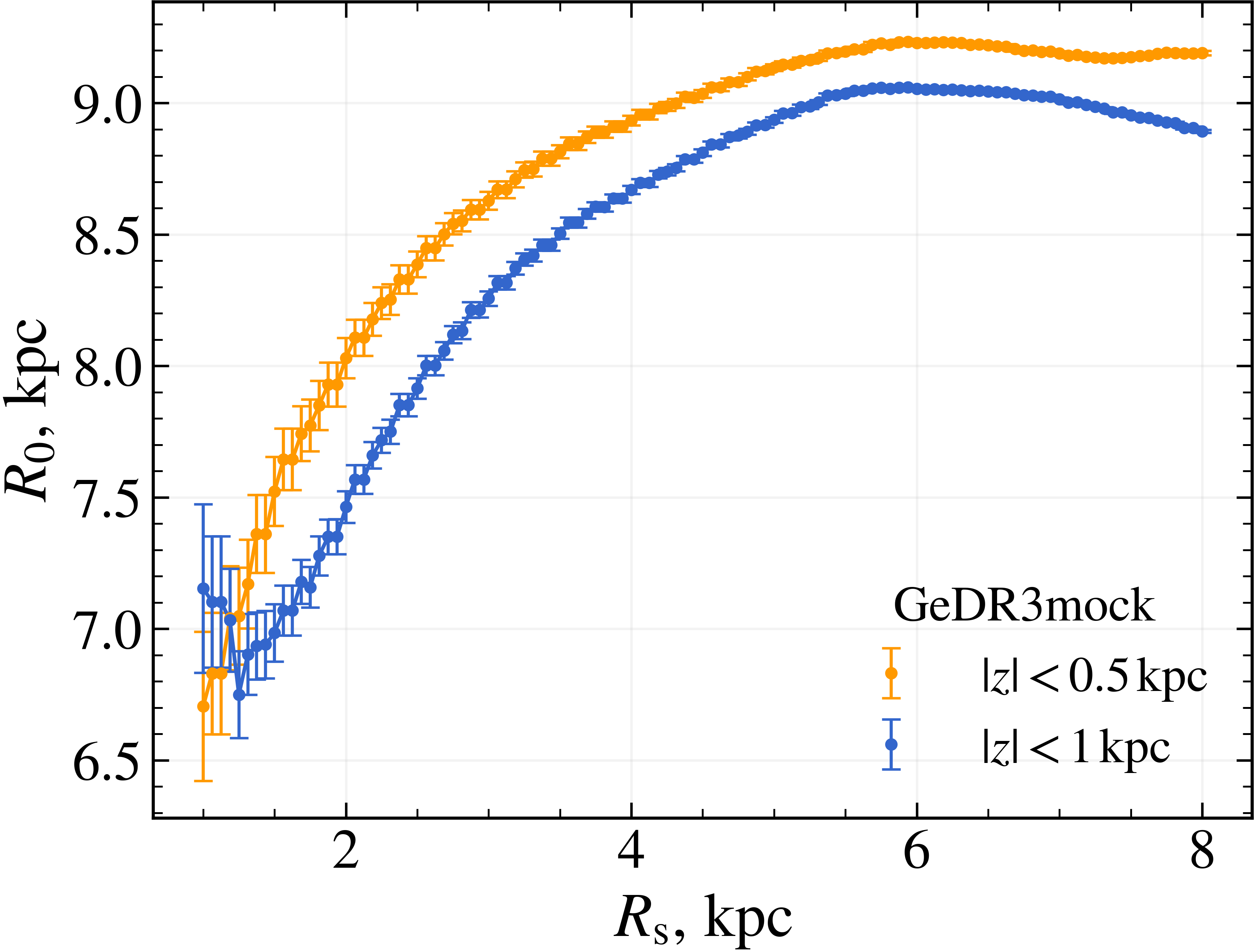}
\hfill
\includegraphics[width=0.49\textwidth]{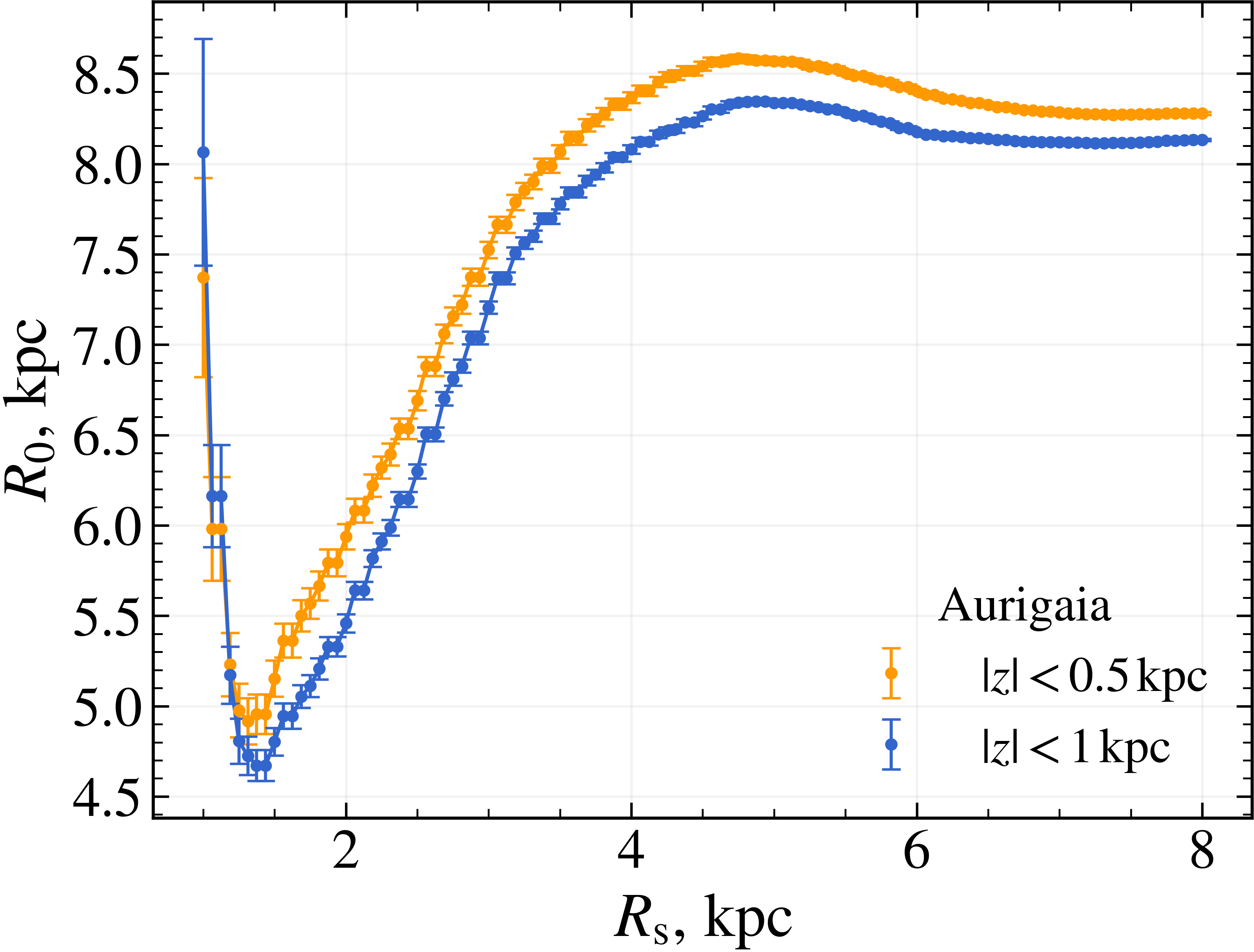}
\caption{Dependence of the derived Galactocentric distance $R_0$ on the heliocentric distance limit $R_s$ for the GeDR3mock (left) and AuriGaia (right) catalogues.}
\label{fig:R0_mock}
\end{figure*}

To investigate the behaviour of the Bottlinger model under controlled conditions, we applied the same fitting procedure to two Gaia-like mock catalogues: GeDR3mock \citep{Rybizki2020} and the ICC mock catalogue based on Auriga halo~6 \citep{Grand2018}. Auriga halo~6 was selected because it represents one of the closest Milky Way analogues within the Auriga simulation suite. It should be emphasized that only the Bottlinger fitting procedure was applied to the mock catalogues; the complete Gaia DR3 analysis pipeline, including the observational selection criteria and Monte Carlo uncertainty propagation, was not reproduced.

The dependence of the derived kinematic parameters on the adopted heliocentric distance limit $R_s$ is presented in Appendix~\ref{app:mocks}. Fig.~\ref{fig:R0_mock} illustrates the corresponding behaviour of the recovered solar Galactocentric distance.

As shown in Fig.~\ref{fig:R0_mock}, the estimated values of $R_0$ gradually converge with increasing heliocentric distance limit and approach an approximately constant value at sufficiently large $R_s$. In contrast to the Gaia DR3 analysis, in which heliocentric distances were obtained by direct inversion of the corrected parallaxes, $r=1/\varpi$, and the stellar sample was restricted to $R_s \leq 5$~kpc because such distance estimates become increasingly unreliable at large distances \citep{BailerJones2023}, the mock catalogues extend to $R_s = 8$~kpc. This allows the behaviour of the $R_0(R_s)$ relation to be examined over a larger spatial volume and demonstrates its convergence towards a stable value.

Table~\ref{tab:gedr3mock} summarizes the kinematic parameters derived for the GeDR3mock catalogue. Since no unique reference values for the Galactic kinematic parameters are available for this mock catalogue, the table is primarily intended to assess the internal consistency of the recovered solutions for the two adopted vertical selections rather than their agreement with a predefined reference model. For consistency with the Gaia DR3 analysis, the final parameter values were determined from the plateau regions of the corresponding $R_s$ dependences using the same MAD-based plateau-selection procedure described in Sections~\ref{sec:solar_motion} and \ref{sec:omega}.

As in the Gaia DR3 analysis, the recovered values of $u_0$, $w_0$, and $\omega_0$ remain nearly unchanged between the two vertical selections, whereas the largest variation is observed for the tangential component of the Solar peculiar velocity, with $v_0$ increasing from $23.35$ to $24.90$~km\,s$^{-1}$. A similar trend is found for the Gaia DR3 sample, where $v_0$ likewise exhibits the largest difference between the two adopted vertical selections, changing from $16.59$ to $16.66$~km\,s$^{-1}$, while the remaining parameters change only marginally. Although the absolute values of $v_0$ differ between the GeDR3mock and Gaia DR3 catalogues, both datasets exhibit the same qualitative behaviour. This suggests that the tangential component is more sensitive to the adopted stellar sample than the other kinematic parameters. This behaviour is also consistent with previous studies, which have shown that $v_0$ is generally the most model-dependent component of the Solar peculiar velocity because of its strong coupling with the adopted Galactic rotation model and the circular velocity of the Local Standard of Rest \citep{Schonrich2010,Schonrich2012,McMillan2017}.

\begin{table}
\centering
\caption{Derived kinematic parameters for the GeDR3mock catalogue.}
\label{tab:gedr3mock}
\begin{tabular}{lcc}
\toprule
Parameter & $|Z|<0.5\,\mathrm{kpc}$ & $|Z|<1.0\,\mathrm{kpc}$ \\
\midrule
$R_0$ (kpc)                        & $9.191  \pm 0.030$ & $9.003  \pm 0.068$ \\
$u_0$ ($\mathrm{km\,s^{-1}}$)      & $10.974 \pm 0.040$ & $10.973 \pm 0.040$ \\
$v_0$ ($\mathrm{km\,s^{-1}}$)      & $23.353 \pm 0.362$ & $24.902 \pm 0.311$ \\
$w_0$ ($\mathrm{km\,s^{-1}}$)      & $7.275  \pm 0.017$ & $7.264  \pm 0.016$ \\
$\omega_0$ ($\mathrm{km\,s^{-1}\,kpc^{-1}}$) 
                                   & $26.912 \pm 0.044$ & $26.680 \pm 0.039$ \\
\bottomrule
\end{tabular}
\end{table}

Unlike GeDR3mock, the AuriGaia catalogue is based on a cosmological simulation with known underlying Galactic parameters. This makes it possible to compare the recovered Bottlinger-model solution directly with the corresponding model values and to evaluate not only the internal consistency of the method but also its ability to recover the principal Galactic kinematic parameters. The corresponding comparison is presented in Table~\ref{tab:aurigaia}.

\begin{table}
\centering
\caption{Input and derived kinematic parameters for the AuriGaia catalogue.}
\label{tab:aurigaia}
\begin{tabular}{lccc}
\toprule
Parameter  & Model & $|Z|<0.5\,\mathrm{kpc}$ & $|Z|<1.0\,\mathrm{kpc}$ \\
\midrule
$R_0$ (kpc)                        & 8.00  & $8.359  \pm 0.126$ & $ 8.123  \pm 0.013$ \\
$u_0$ ($\mathrm{km\,s^{-1}}$)      & 11.10 & $18.513 \pm 0.383$ & $ 18.373 \pm 0.507$ \\
$v_0$ ($\mathrm{km\,s^{-1}}$)      & 12.24 & $21.549 \pm 0.688$ & $ 24.204 \pm 0.736$ \\
$w_0$ ($\mathrm{km\,s^{-1}}$)      & 7.25  & $6.428  \pm 0.175$ & $ 6.357  \pm 0.264$ \\
$\omega_0$ ($\mathrm{km\,s^{-1}\,kpc^{-1}}$)
                                   & 28.09 & $27.929 \pm 0.209$ & $ 27.292 \pm 0.163$ \\
\bottomrule
\end{tabular}
\end{table}

For the extended sample ($|z|<1.0$~kpc), the recovered solar Galactocentric distance is $R_0=8.123\pm0.013$~kpc, differing by only about $1.5\%$ from the adopted model value. In contrast, noticeably larger discrepancies are found for several of the remaining kinematic parameters, particularly the components of the Solar peculiar motion. This behaviour is expected because the stellar velocity field in a cosmological simulation is considerably more complex than the axisymmetric differential-rotation model described by the Bottlinger equations and contains non-axisymmetric structures, streaming motions, spiral-arm perturbations, and departures from purely circular orbits. Consequently, the recovered Solar-motion components effectively absorb part of these non-axisymmetric motions and should not be interpreted as direct counterparts of the input model values. Nevertheless, the close agreement between the recovered and input values of $R_0$ demonstrates that the Bottlinger-based approach remains capable of reliably constraining the fundamental geometric scale of the Galaxy even when applied to a realistic cosmological stellar system.

Overall, the experiments with Gaia-like mock catalogues demonstrate that the adopted Bottlinger-based approach provides stable estimates of the Galactocentric distance and reproduces the large-scale Galactic velocity field with good accuracy. At the same time, the tests confirm that the tangential component of the Solar peculiar motion is the most sensitive parameter to the adopted stellar sample, whereas the remaining kinematic parameters remain comparatively stable. These results support the robustness of the proposed methodology when applied to large astrometric datasets such as Gaia DR3.

\section{Conclusions}
\label{sec:con}

In this work, we applied the Bottlinger kinematic model to a sample of approximately 13 million Gaia DR3 stars with full six-dimensional phase-space information. To investigate the influence of the vertical structure of the Galactic disc, the analysis was carried out independently for two subsamples with $|z|<0.5$ and $|z|<1.0$~kpc.

Table~\ref{tab:results} summarizes the final Galactic kinematic parameters derived for the two subsamples. 

\begin{table}
\centering
\caption{Final Galactic kinematic parameters derived for the two vertical subsamples.}
\label{tab:results}
\begin{tabular}{lcc}
\toprule
Parameter & $|z|<0.5\,\mathrm{kpc}$ & $|z|<1.0\,\mathrm{kpc}$ \\
\midrule
$R_0$ (kpc)                         & $8.165 \pm 0.024$ & $8.333 \pm 0.022$ \\
$u_0$ ($\mathrm{km\,s^{-1}}$)       & $9.261 \pm 0.006$ & $9.264 \pm 0.006$ \\
$v_0$ ($\mathrm{km\,s^{-1}}$)       & $16.593\pm 0.008$ & $16.663\pm 0.004$ \\
$w_0$ ($\mathrm{km\,s^{-1}}$)       & $7.686 \pm 0.008$ & $7.697 \pm 0.008$ \\
$\omega_0$ ($\mathrm{km\,s^{-1}\,kpc^{-1}}$)
                                    & $28.477\pm 0.002$ & $28.488\pm 0.001$ \\
\bottomrule
\end{tabular}
\end{table}

The comparison shows that the Solar peculiar velocity and the local angular velocity are remarkably stable with respect to the adopted vertical selection. In contrast, the inferred solar Galactocentric distance exhibits a systematic increase from $8.165$ to $8.333$~kpc when the vertical extent of the sample is increased from $|z|<0.5$ to $|z|<1.0$~kpc. This result demonstrates that, although the local kinematic parameters are relatively robust against moderate changes in the tracer population, the determination of $R_0$ remains sensitive to the vertical structure and spatial distribution of the adopted stellar sample.

Overall, the derived Galactic kinematic parameters are consistent with recent independent determinations based on a wide range of observational tracers and analysis techniques. Our results demonstrate that the Bottlinger model, when applied to high-quality Gaia DR3 data, provides a stable and self-consistent description of the local Galactic kinematics. At the same time, the comparison of different vertical subsamples shows that the properties of the tracer population should be taken into account when interpreting the inferred solar Galactocentric distance.

\section*{Acknowledgements}
\label{sec:acknowledgements}
This work has made use of data from the European Space Agency (ESA) mission {\it Gaia} (\url{https://www.cosmos.esa.int/gaia}), processed by the {\it Gaia} Data Processing and Analysis Consortium (DPAC, \url{https://www.cosmos.esa.int/web/gaia/dpac/consortium}). Funding for the DPAC has been provided by national institutions, in particular the institutions participating in the {\it Gaia} Multilateral Agreement. 

\section*{Data Availability}
The Gaia DR3 catalogue used in this work is publicly available in electronic form through the Gaia Archive or CDS via anonymous FTP at cdsarc.u-strasbg.fr (130.79.128.5). The software code used in this paper can be made available on personal request by e-mail: \href{mailto:odynets.d.v@gmail.com} {odynets.d.v@gmail.com} or \href{mailto:akhmetovvs@gmail.com} {akhmetovvs@gmail.com}.

\bibliographystyle{mnras}
\bibliography{thebib}

\appendix

\section{Gaia-like mock catalogues analysis}
\label{app:mocks}

This appendix presents the complete set of kinematic parameter dependences obtained for the simulated GeDR3mock and AuriGaia catalogues. The upper panels show the results for GeDR3mock, whereas the lower panels correspond to the AuriGaia catalogue.

\begin{figure*}
\centering
\includegraphics[width=0.33\textwidth]{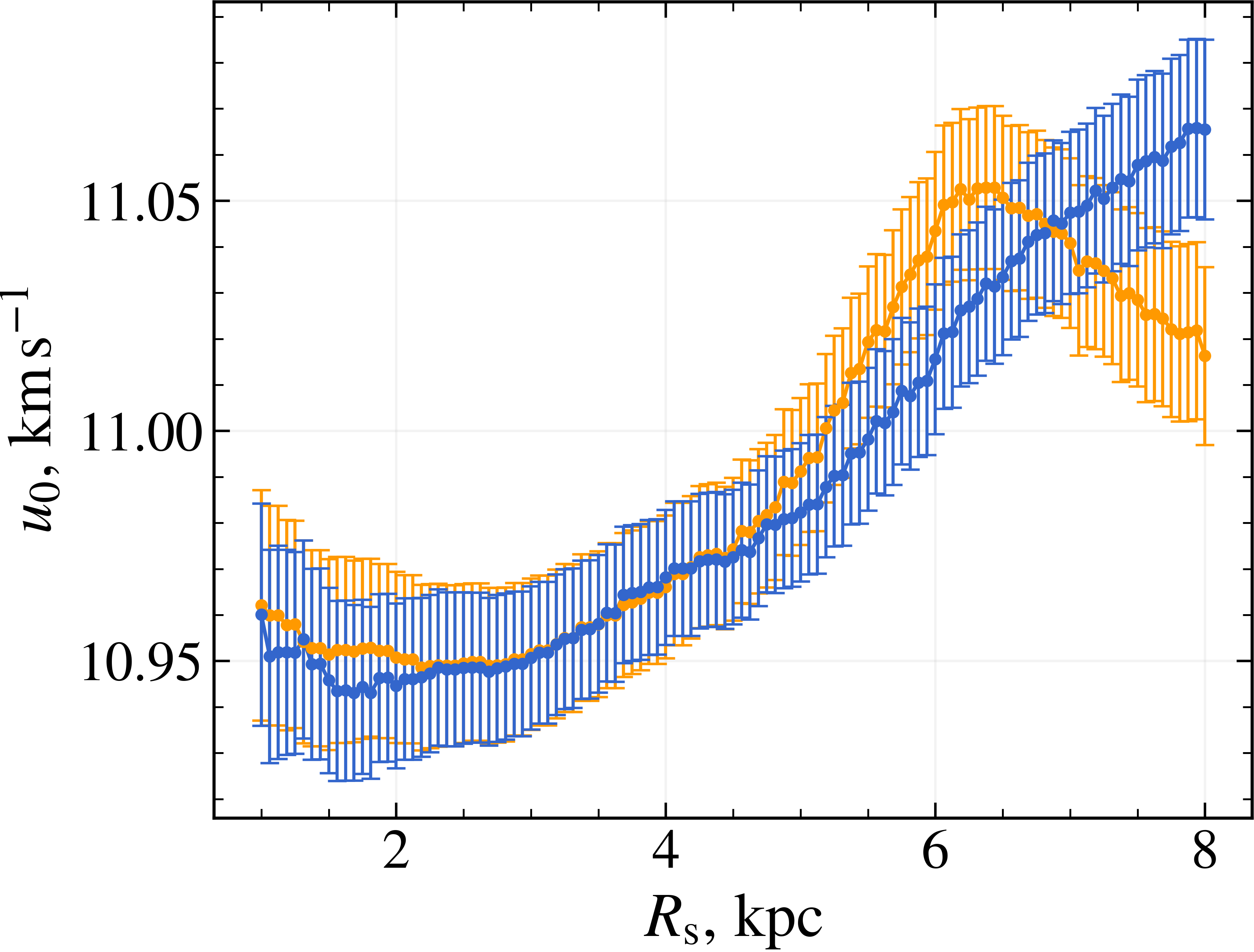}
\hfill
\includegraphics[width=0.33\textwidth]{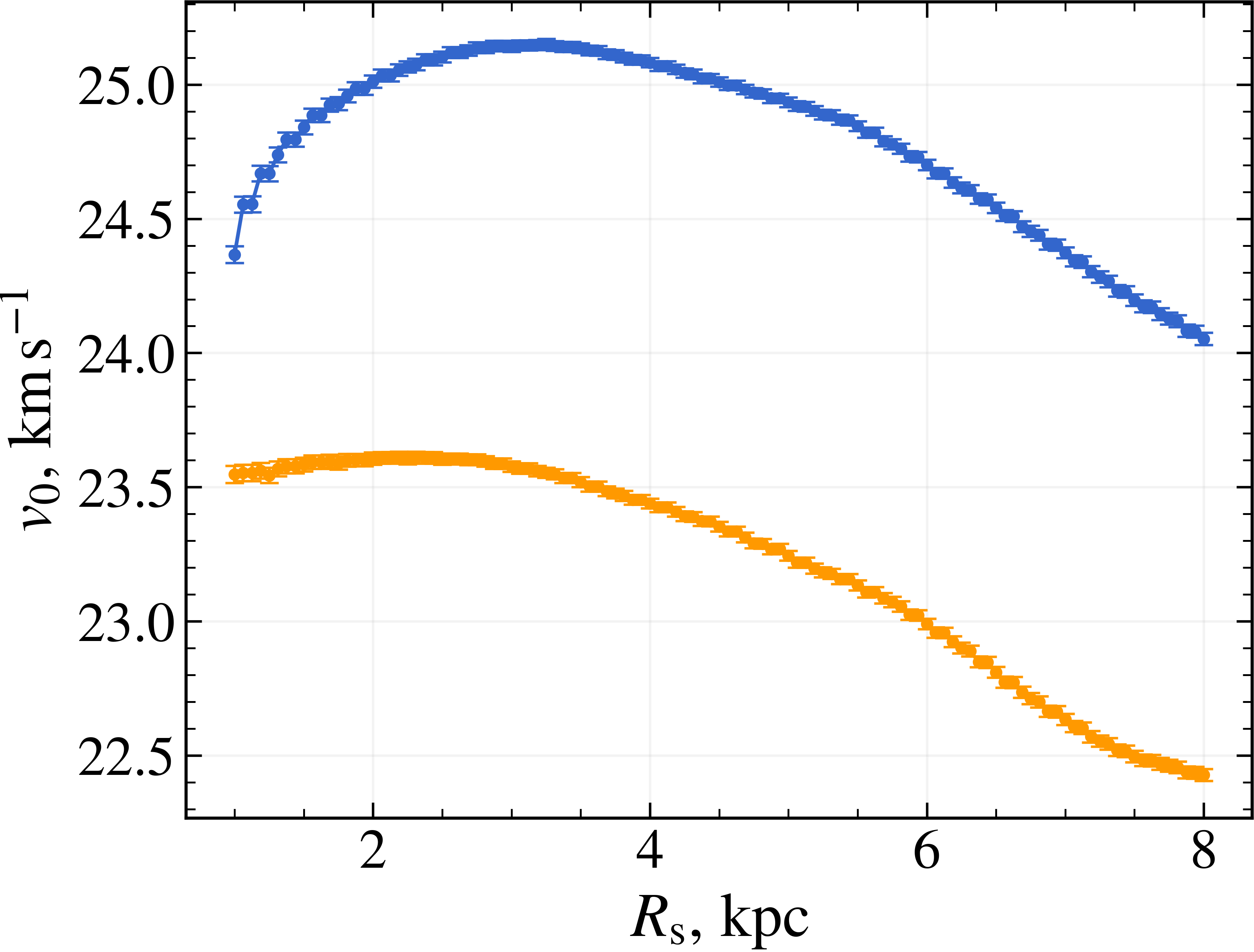}
\hfill
\includegraphics[width=0.33\textwidth]{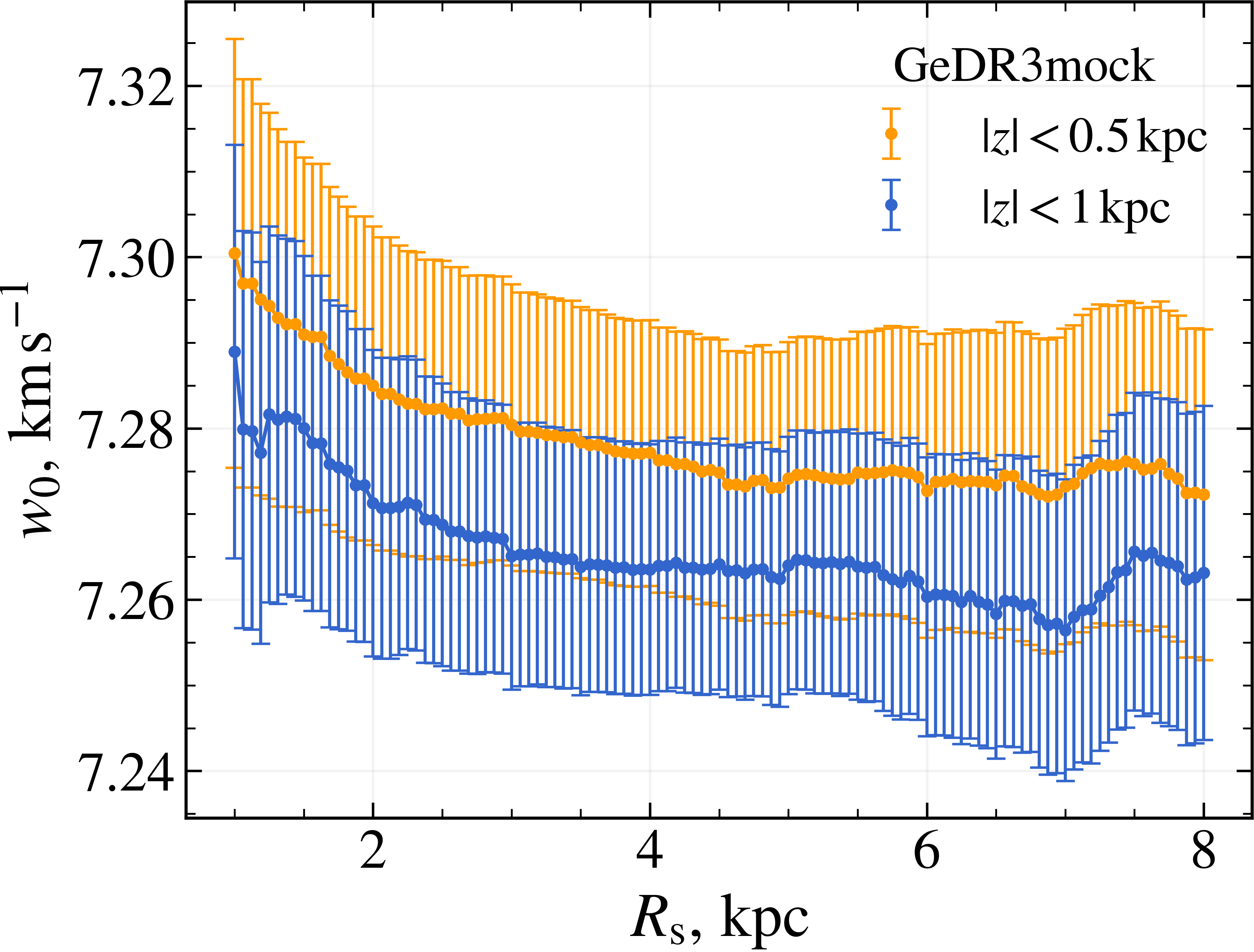}
\vspace{0.5em}
\includegraphics[width=0.33\textwidth]{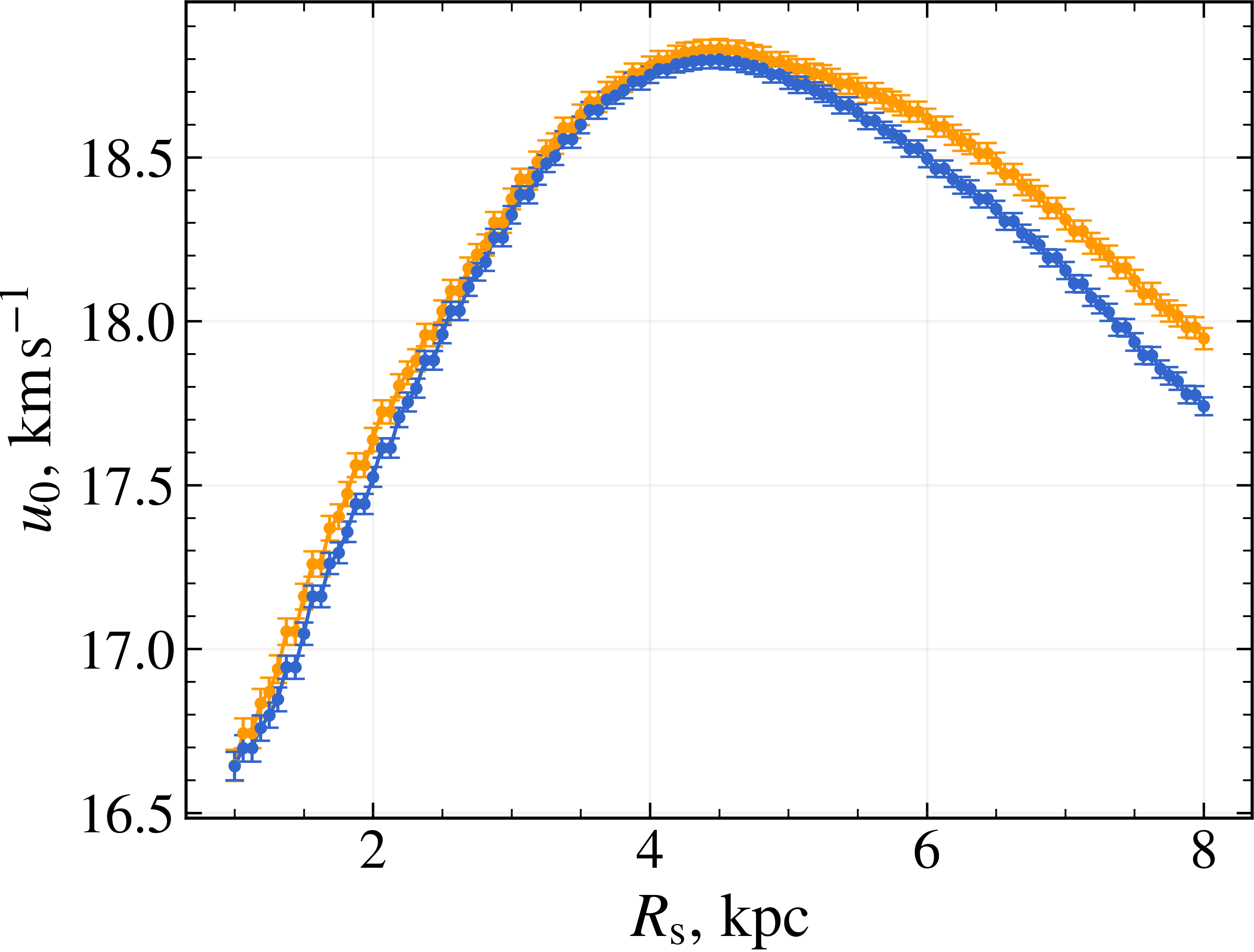}
\hfill
\includegraphics[width=0.33\textwidth]{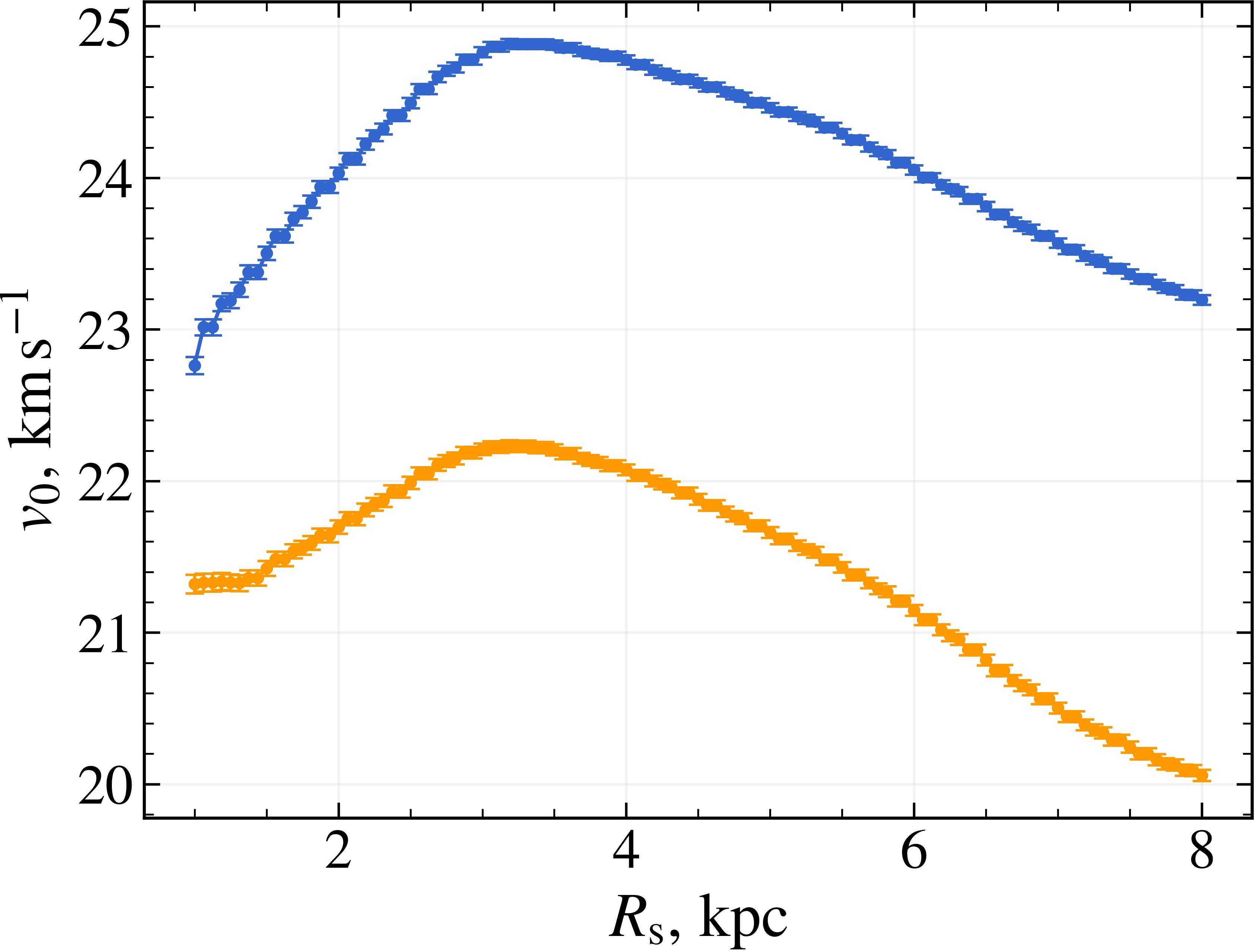}
\hfill
\includegraphics[width=0.33\textwidth]{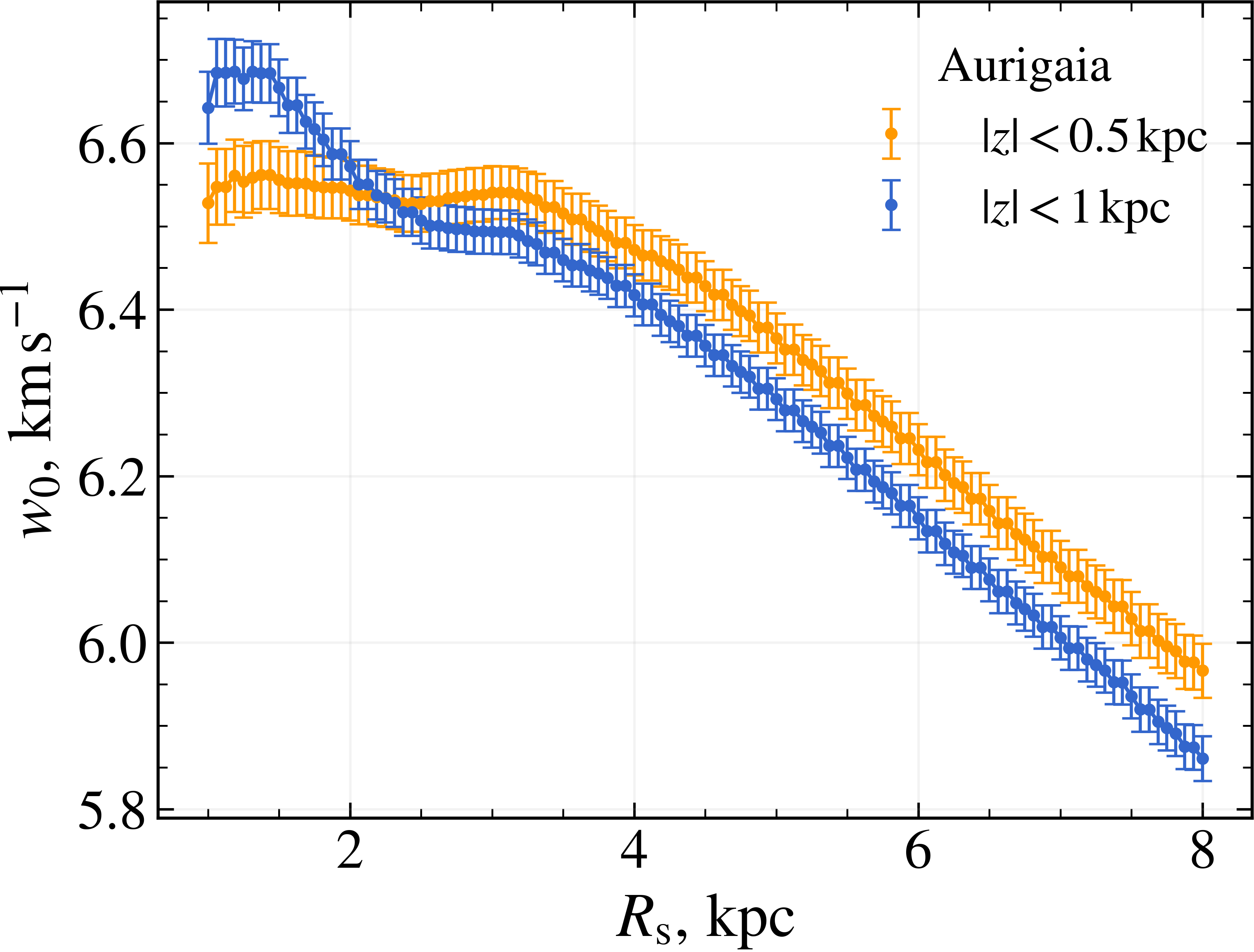}

\caption{Dependence of the recovered Solar peculiar motion components ($u_0$, $v_0$, and $w_0$) on the adopted heliocentric distance limit for the GeDR3mock (upper row) and AuriGaia (lower row) catalogues.}
\label{fig:solar_motion_mock}
\end{figure*}

\begin{figure*}
\centering
\includegraphics[width=0.33\textwidth]{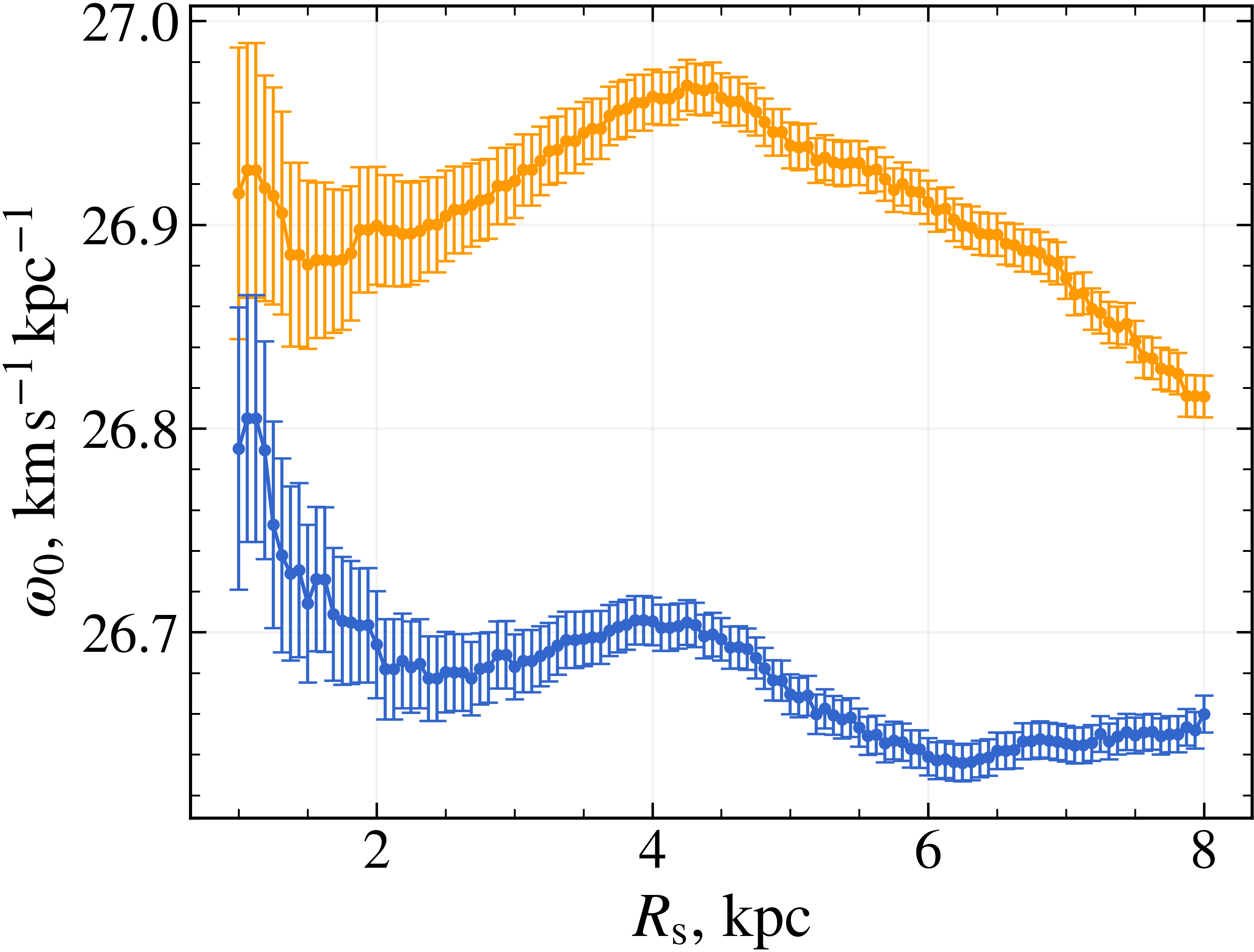}
\hfill
\includegraphics[width=0.33\textwidth]{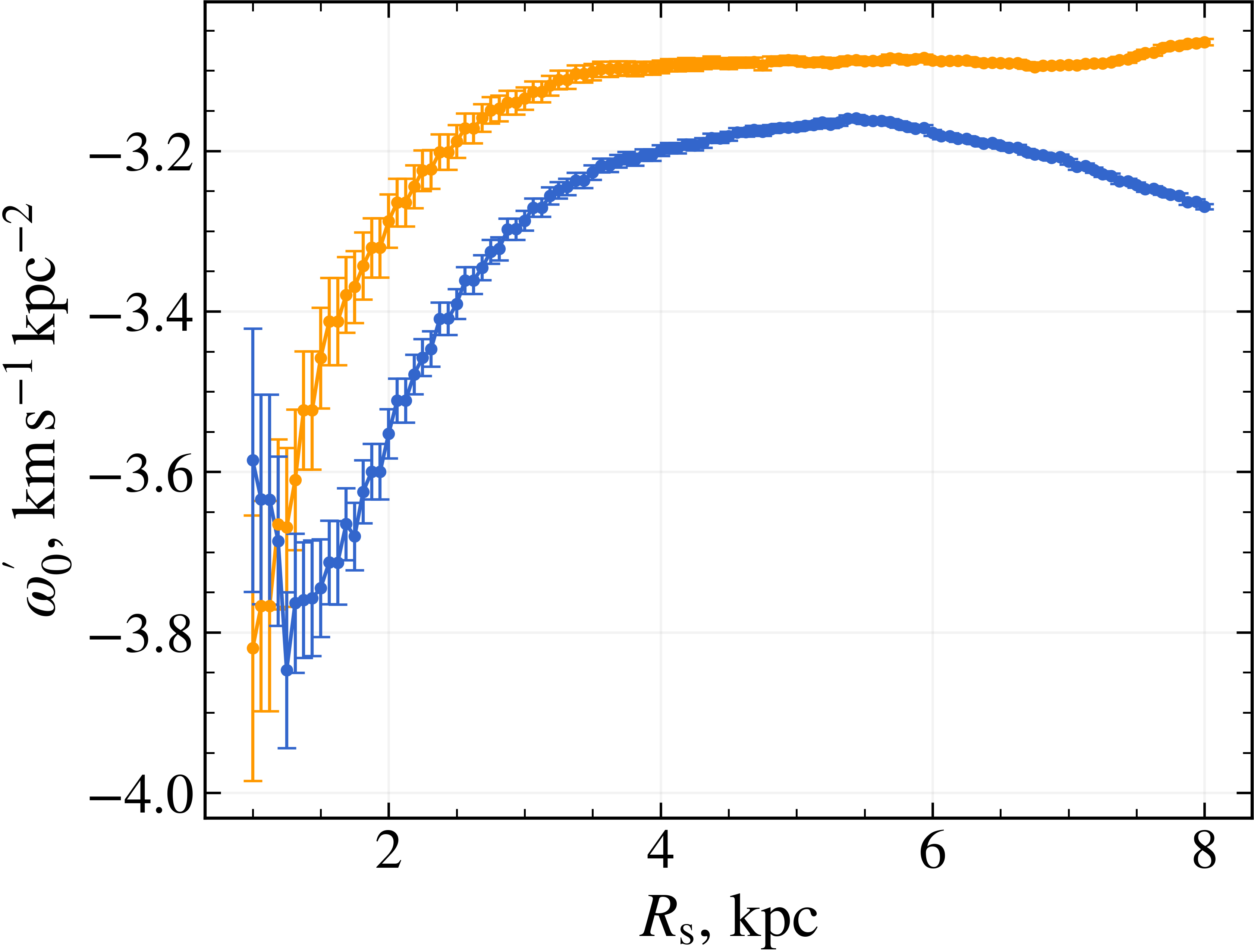}
\hfill
\includegraphics[width=0.33\textwidth]{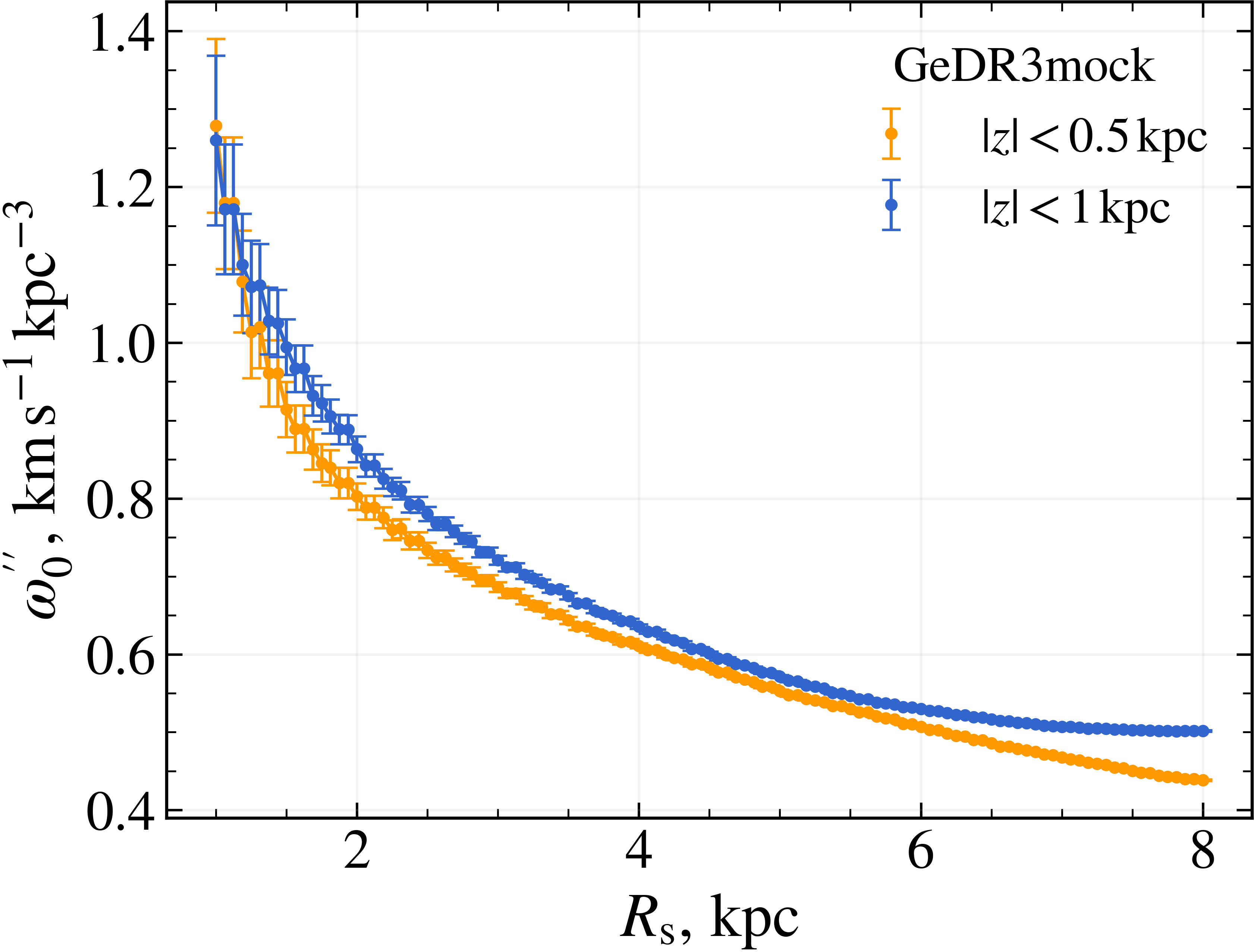}
\vspace{0.5em}
\includegraphics[width=0.33\textwidth]{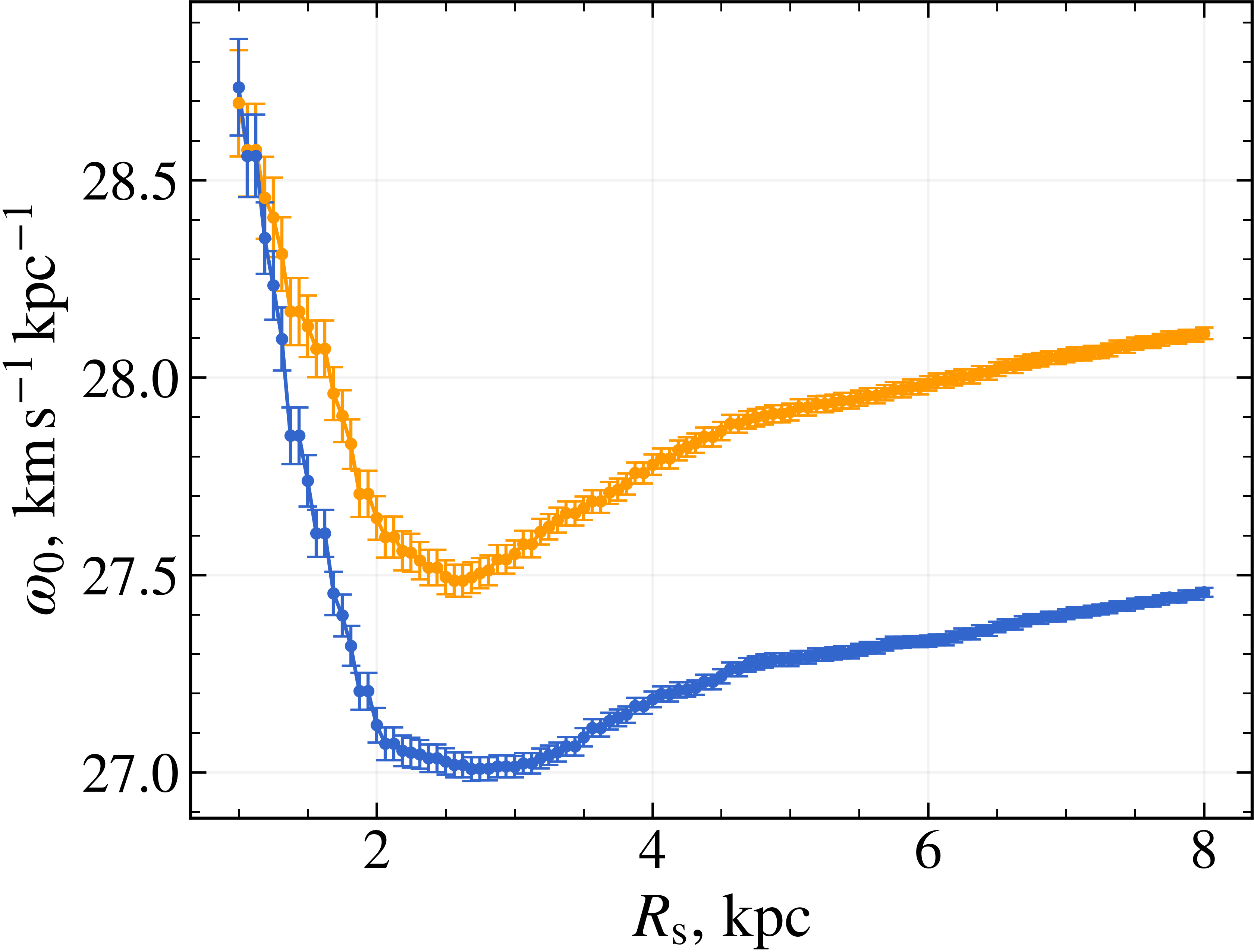}
\hfill
\includegraphics[width=0.33\textwidth]{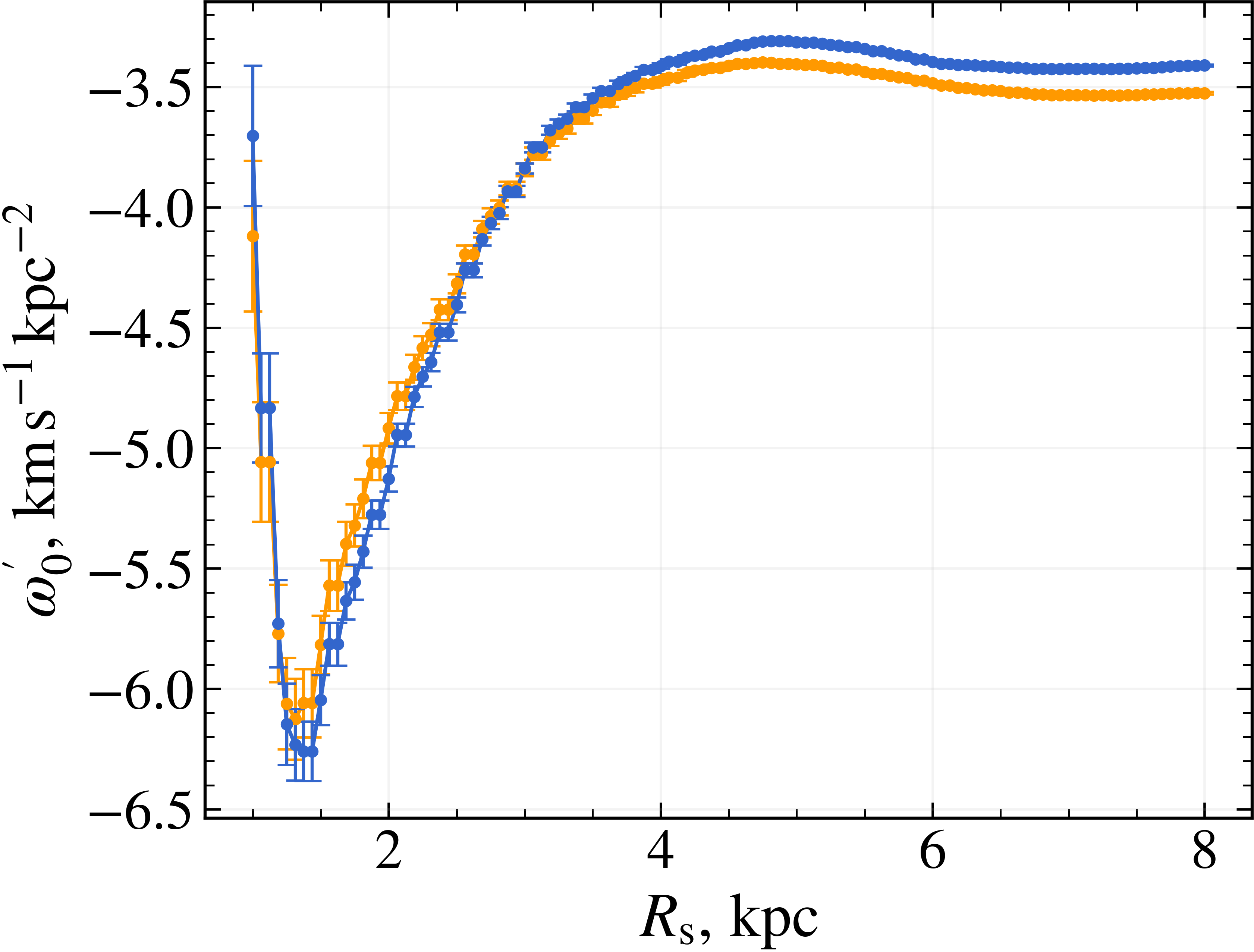}
\hfill
\includegraphics[width=0.33\textwidth]{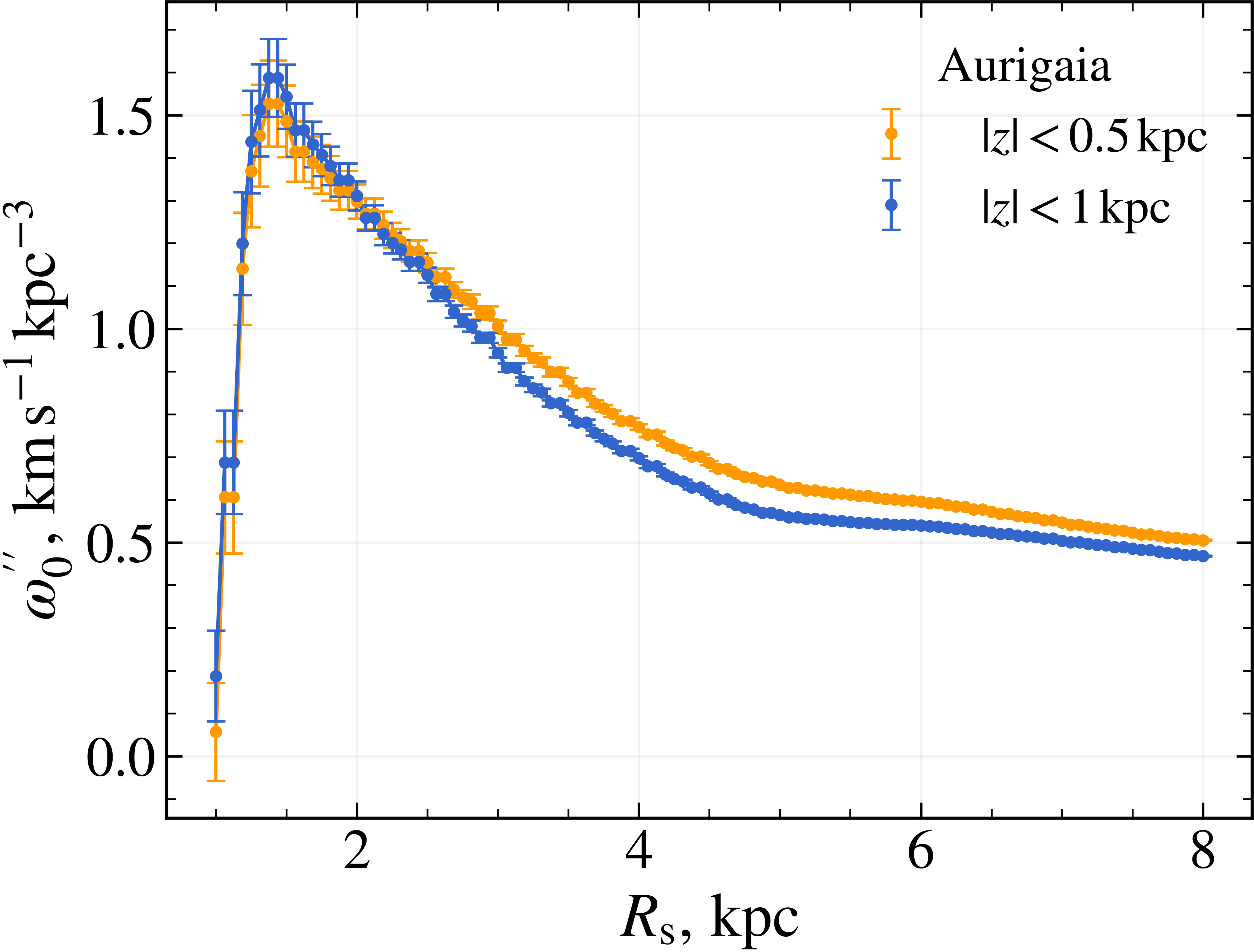}

\caption{Dependence of the recovered local Galactic angular velocity ($\omega_0$) and its first two radial derivatives ($\omega'_0$ and $\omega''_0$) on the adopted heliocentric distance limit for the GeDR3mock (upper row) and AuriGaia (lower row) catalogues.}
\label{fig:omega_mock}
\end{figure*}

\begin{figure*}
\centering
\includegraphics[width=0.49\textwidth]{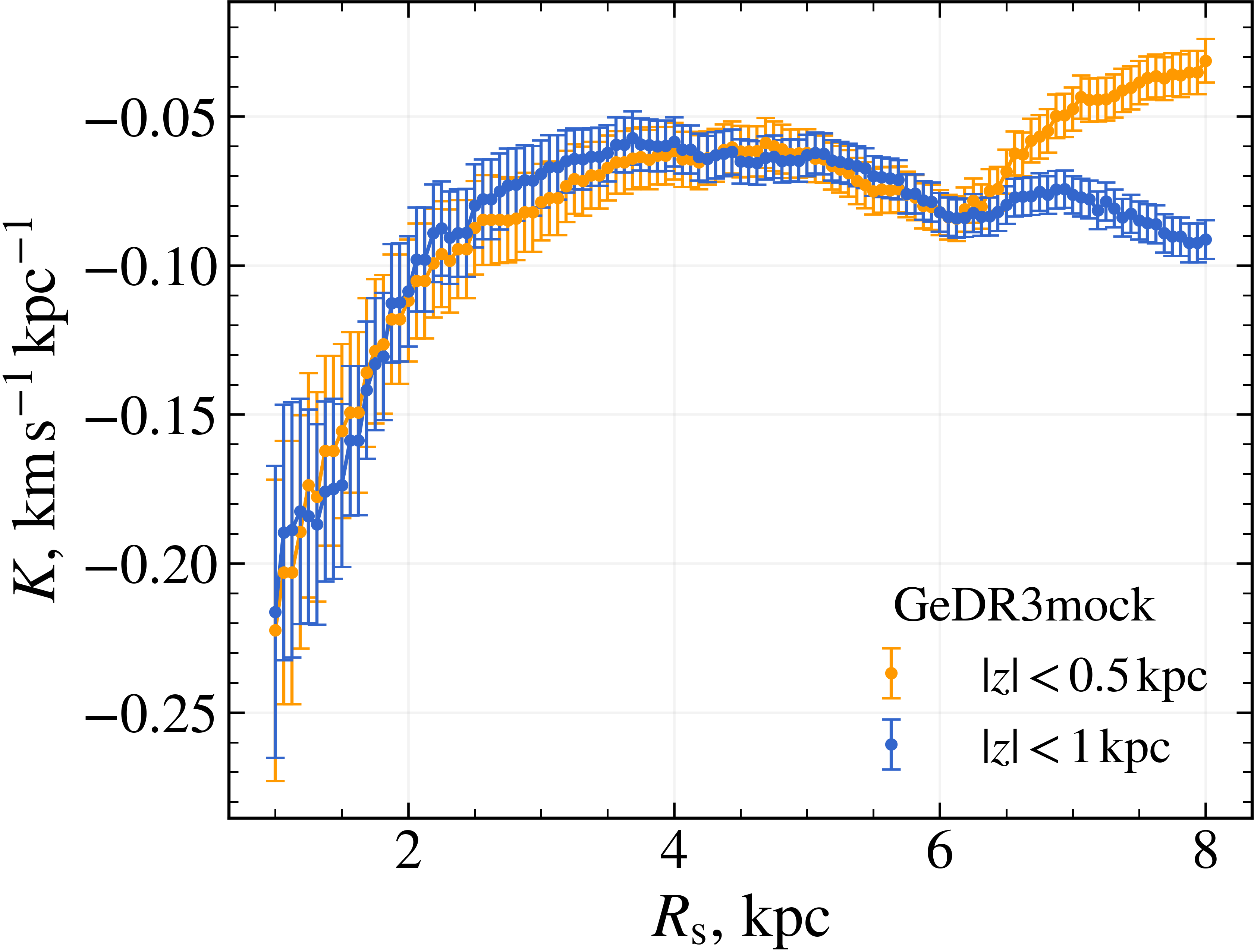}
\hfill
\includegraphics[width=0.49\textwidth]{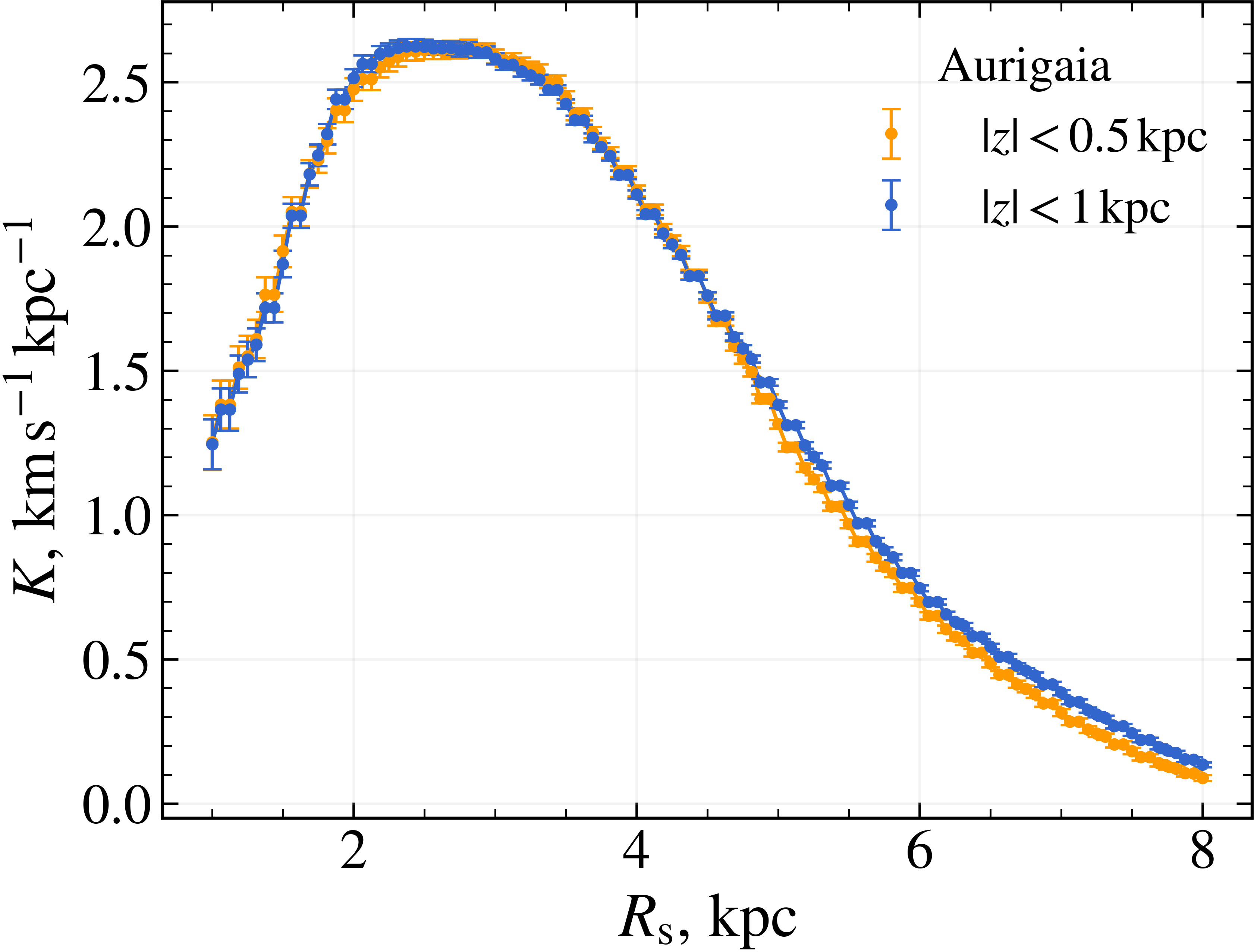}
\caption{Dependence of the radial expansion parameter $K$ on the adopted heliocentric distance limit for the GeDR3mock (left) and AuriGaia (right) catalogues.}
\label{fig:K_mock}
\end{figure*}

\bsp
\label{lastpage}
\end{document}